\documentclass[a4paper,12pt]{article}

\usepackage[top=1.2in,right=0.5in,left=.5in,bottom=1.2in]{geometry}

\usepackage{latexsym}
\usepackage{amssymb}
\usepackage{float}
\usepackage{graphics,graphicx}
\usepackage{tabularx}
\usepackage{amsfonts}
\usepackage{amsmath}
\usepackage{enumerate}
\usepackage{booktabs}
\usepackage{multirow}
\usepackage{sectsty}
\usepackage[affil-it,auth-sc]{authblk}
\usepackage{subfig}
\usepackage{color}
\usepackage{mathtools}
\usepackage{mathrsfs}
\usepackage{bmpsize}
\usepackage{xcolor}
\usepackage{siunitx}
\usepackage{hyperref}

\sectionfont{\normalsize}
\subsectionfont{\normalsize}
\usepackage{fancyhdr}            
\fancypagestyle{plain}{%
  \fancyhead{}                                     
  \fancyfoot[CE,CO]{\thepage}       
  \fancyhead[CE,CO]{\textcolor[rgb]{0.64,0.15,0.15}{Published in \emph{Philosophical Transactions of the Royal Society A} (2019) \textbf{377}: 20190101
  \\
doi: https://doi.org/10.1098/rsta.2019.0101}}
\setlength{\headheight}{14.5pt}}

\newcommand{\footmsg}[1]{%
  \let\temp\thempfn%
  \def\thempfs{}
  \footnotetext{#1}
  \let\tempfn\temp}

\begin{document}

\newcommand{\singlespace}{\baselineskip=12pt\lineskiplimit=0pt\lineskip=0pt}
\def\ds{\displaystyle}

\newcommand{\beq}{\begin{equation}}
\newcommand{\eeq}{\end{equation}}
\newcommand{\lb}{\label}
\newcommand{\ph}{\phantom}
\newcommand{\beqar}{\begin{eqnarray}}
\newcommand{\eeqar}{\end{eqnarray}}
\newcommand{\barr}{\begin{array}}
\newcommand{\earr}{\end{array}}
\newcommand{\jump}{\parallel}
\newcommand{\Ehat}{\hat{E}}
\newcommand{\That}{\hat{\bf T}}
\newcommand{\Ahat}{\hat{A}}
\newcommand{\chat}{\hat{c}}
\newcommand{\shat}{\hat{s}}
\newcommand{\khat}{\hat{k}}
\newcommand{\muhat}{\hat{\mu}}
\newcommand{\mc}{M^{\scriptscriptstyle C}}
\newcommand{\mei}{M^{\scriptscriptstyle M,EI}}
\newcommand{\mec}{M^{\scriptscriptstyle M,EC}}
\newcommand{\hbeta}{{\hat{\beta}}}
\newcommand{\rec}[2]{\left( #1 #2 \ds{\frac{1}{#1}}\right)}
\newcommand{\rep}[2]{\left( {#1}^2 #2 \ds{\frac{1}{{#1}^2}}\right)}
\newcommand{\derp}[2]{\ds{\frac {\partial #1}{\partial #2}}}
\newcommand{\derpn}[3]{\ds{\frac {\partial^{#3}#1}{\partial #2^{#3}}}}
\newcommand{\dert}[2]{\ds{\frac {d #1}{d #2}}}
\newcommand{\dertn}[3]{\ds{\frac {d^{#3} #1}{d #2^{#3}}}}
\newcommand{\ct}{\captionof{table}}
\newcommand{\cf}{\captionof{figure}}

\def\c{{\circ}}
\def\bob{{\, \underline{\overline{\otimes}} \,}}
\def\ob{{\, \underline{\otimes} \,}}
\def\scalp{\mbox{\boldmath$\, \cdot \, $}}
\def\gdp{\makebox{\raisebox{-.215ex}{$\Box$}\hspace{-.778em}$\times$}}
\def\daa{\makebox{\raisebox{-.050ex}{$-$}\hspace{-.550em}$: ~$}}
\def\mK{\mbox{${\mathcal{K}}$}}
\def\cK{\mbox{${\mathbb {K}}$}}

\def\Xint#1{\mathchoice
   {\XXint\displaystyle\textstyle{#1}}%
   {\XXint\textstyle\scriptstyle{#1}}%
   {\XXint\scriptstyle\scriptscriptstyle{#1}}%
   {\XXint\scriptscriptstyle\scriptscriptstyle{#1}}%
   \!\int}
\def\XXint#1#2#3{{\setbox0=\hbox{$#1{#2#3}{\int}$}
     \vcenter{\hbox{$#2#3$}}\kern-.5\wd0}}
\def\ddashint{\Xint=}
\def\fpint{\Xint=}
\def\dashint{\Xint-}
\def\cpvint{\Xint-}
\def\intl{\int\limits}
\def\cpvintl{\cpvint\limits}
\def\fpintl{\fpint\limits}
\def\ointl{\oint\limits}
\def\bA{{\bf A}}
\def\ba{{\bf a}}
\def\bB{{\bf B}}
\def\bb{{\bf b}}
\def\bc{{\bf c}}
\def\bC{{\bf C}}
\def\bD{{\bf D}}
\def\bE{{\bf E}}
\def\be{{\bf e}}
\def\bbf{{\bf f}}
\def\bF{{\bf F}}
\def\bG{{\bf G}}
\def\bg{{\bf g}}
\def\bi{{\bf i}}
\def\bH{{\bf H}}
\def\bK{{\bf K}}
\def\bL{{\bf L}}
\def\bM{{\bf M}}
\def\bN{{\bf N}}
\def\bn{{\bf n}}
\def\bm{{\bf m}}
\def\b0{{\bf 0}}
\def\bo{{\bf o}}
\def\bX{{\bf X}}
\def\bx{{\bf x}}
\def\bP{{\bf P}}
\def\bp{{\bf p}}
\def\bQ{{\bf Q}}
\def\bq{{\bf q}}
\def\bR{{\bf R}}
\def\bS{{\bf S}}
\def\bs{{\bf s}}
\def\bT{{\bf T}}
\def\bt{{\bf t}}
\def\bU{{\bf U}}
\def\bu{{\bf u}}
\def\bv{{\bf v}}
\def\bw{{\bf w}}
\def\bW{{\bf W}}
\def\by{{\bf y}}
\def\bz{{\bf z}}
\def\T{{\bf T}}
\def\Te{\textrm{T}}
\def\Id{{\bf I}}
\def\bxi{\mbox{\boldmath${\xi}$}}
\def\balpha{\mbox{\boldmath${\alpha}$}}
\def\bbeta{\mbox{\boldmath${\beta}$}}
\def\bepsilon{\mbox{\boldmath${\epsilon}$}}
\def\bvarepsilon{\mbox{\boldmath${\varepsilon}$}}
\def\bomega{\mbox{\boldmath${\omega}$}}
\def\bphi{\mbox{\boldmath${\phi}$}}
\def\bsigma{\mbox{\boldmath${\sigma}$}}
\def\bfeta{\mbox{\boldmath${\eta}$}}
\def\bDelta{\mbox{\boldmath${\Delta}$}}
\def\btau{\mbox{\boldmath $\tau$}}
\def\tr{{\rm tr}}
\def\dev{{\rm dev}}
\def\div{{\rm div}}
\def\Div{{\rm Div}}
\def\Grad{{\rm Grad}}
\def\grad{{\rm grad}}
\def\Lin{{\rm Lin}}
\def\Sym{{\rm Sym}}
\def\Skw{{\rm Skew}}
\def\abs{{\rm abs}}
\def\Re{{\rm Re}}
\def\Im{{\rm Im}}
\def\capB{\mbox{\boldmath${\mathsf B}$}}
\def\capC{\mbox{\boldmath${\mathsf C}$}}
\def\capD{\mbox{\boldmath${\mathsf D}$}}
\def\capE{\mbox{\boldmath${\mathsf E}$}}
\def\capG{\mbox{\boldmath${\mathsf G}$}}
\def\tcapG{\tilde{\capG}}
\def\capH{\mbox{\boldmath${\mathsf H}$}}
\def\capK{\mbox{\boldmath${\mathsf K}$}}
\def\capL{\mbox{\boldmath${\mathsf L}$}}
\def\capM{\mbox{\boldmath${\mathsf M}$}}
\def\capR{\mbox{\boldmath${\mathsf R}$}}
\def\capW{\mbox{\boldmath${\mathsf W}$}}

\def\i{\mbox{${\mathrm i}$}}
\def\mC{\mbox{\boldmath${\mathcal C}$}}
\def\mB{\mbox{${\mathcal B}$}}
\def\mE{\mbox{${\mathcal{E}}$}}
\def\mL{\mbox{${\mathcal{L}}$}}
\def\mK{\mbox{${\mathcal{K}}$}}
\def\mV{\mbox{${\mathcal{V}}$}}
\def\C{\mbox{\boldmath${\mathcal C}$}}
\def\E{\mbox{\boldmath${\mathcal E}$}}

\def\AAM{{\it Advances in Applied Mechanics }}
\def\ACME{{\it Arch. Comput. Meth. Engng.}}
\def\ARMA{{\it Arch. Rat. Mech. Analysis}}
\def\AMR{{\it Appl. Mech. Rev.}}
\def\ASCEEM{{\it ASCE J. Eng. Mech.}}
\def\ACTA{{\it Acta Mater.}}
\def\CMAME {{\it Comput. Meth. Appl. Mech. Engrg.}}
\def\CRAS{{\it C. R. Acad. Sci. Paris}}
\def\CRM{{\it Comptes Rendus M\'ecanique}}
\def\EFM{{\it Eng. Fracture Mechanics}}
\def\EJMA{{\it Eur.~J.~Mechanics-A/Solids}}
\def\IJES{{\it Int. J. Eng. Sci.}}
\def\IJF{{\it Int. J. Fracture}}
\def\IJMS{{\it Int. J. Mech. Sci.}}
\def\IJNAMG{{\it Int. J. Numer. Anal. Meth. Geomech.}}
\def\IJP{{\it Int. J. Plasticity}}
\def\IJSS{{\it Int. J. Solids Structures}}
\def\IngA{{\it Ing. Archiv}}
\def\JAM{{\it J. Appl. Mech.}}
\def\JAP{{\it J. Appl. Phys.}}
\def\JAE{{\it J. Aerospace Eng.}}
\def\JE{{\it J. Elasticity}}
\def\JM{{\it J. de M\'ecanique}}
\def\JMPS{{\it J. Mech. Phys. Solids}}
\def\JSV{{\it J. Sound and Vibration}}
\def\MACRO{{\it Macromolecules}}
\def\MMT{{\it Mech. Mach. Th.}}
\def\MOM{{\it Mech. Materials}}
\def\MMS{{\it Math. Mech. Solids}}
\def\MMT{{\it Metall. Mater. Trans. A}}
\def\MPCPS{{\it Math. Proc. Camb. Phil. Soc.}}
\def\MSE{{\it Mater. Sci. Eng.}}
\def\NATURE{{\it Nature}}
\def\NATUREM{{\it Nature Mater.}}
\def\PHIL{{\it Phil. Trans. R. Soc.}}
\def\PMPS{{\it Proc. Math. Phys. Soc.}}
\def\PNAS{{\it Proc. Nat. Acad. Sci.}}
\def\PRE{{\it Phys. Rev. E}}
\def\PRL{{\it Phys. Rev. Letters}}
\def\PRSL{{\it Proc. R. Soc.}}
\def\RIIT{{\it Rozprawy Inzynierskie - Engineering Transactions}}
\def\ROCK{{\it Rock Mech. and Rock Eng.}}
\def\QAM{{\it Quart. Appl. Math.}}
\def\QJMAM{{\it Quart. J. Mech. Appl. Math.}}
\def\SCIENCE{{\it Science}}
\def\SCRMAT{{\it Scripta Mater.}}
\def\SM{{\it Scripta Metall.}}
\def\ZAMM{{\it Z. Angew. Math. Mech.}}
\def\ZAMP{{\it Z. Angew. Math. Phys.}}
\def\ZVDI{{\it Z. Verein. Deut. Ing.}}

\def\salto#1#2{
[\mbox{\hspace{-#1em}}[#2]\mbox{\hspace{-#1em}}]}

\renewcommand\Affilfont{\itshape}
\setlength{\affilsep}{1em}
\renewcommand\Authsep{, }
\renewcommand\Authand{ and }
\renewcommand\Authands{ and }
\setcounter{Maxaffil}{2}

\title{Nested Bloch waves in elastic structures \\
with configurational forces}

\author{F. Dal Corso$^{\text{A}}$, D. Tallarico$^\text{B}$, N.V. Movchan$^\text{C}$, \\
A.B. Movchan$^\text{C}$, and D. Bigoni$^\text{A}$}
 \affil[A]{DICAM, University of Trento, via~Mesiano~77, I-38123
Trento, Italy.}
 \affil[B]{EMPA - Swiss Federal Laboratories for Materials Science and Technology, Laboratory for Acoustics/Noise Control, D{\"u}bendorf, Switzerland.}
 \affil[C]{Department of Mathematical 
	Sciences,
	Mathematical Sciences Building, The University of Liverpool, L69 7ZL, Liverpool, United Kingdom
	}

\date{}
\maketitle \footnotetext[1]{Corresponding author: Davide Bigoni fax:
+39 0461 282599; tel.: +39 0461 282507; web-site:
http://www.ing.unitn.it/$\sim$bigoni/; e-mail: bigoni@ing.unitn.it.}

\date{}
\maketitle

\begin{abstract}
Small axial and flexural oscillations are analyzed for a periodic and infinite structure, constrained by sliding sleeves and composed of  elastic beams.  A nested Bloch-Floquet technique is introduced to treat the non-linear coupling between longitudinal and transverse displacements induced by the configurational forces generated at the sliding sleeve ends. The action of configurational forces is shown to play
an important role from two perspectives. First, the band gap structure for purely longitudinal vibration is broken so that axial propagation may occur at  frequencies that are forbidden in the absence of a transverse oscillation and, second, a flexural oscillation may induce axial resonance, a situation in which the longitudinal vibrations tend to become unbounded.
The presented results disclose the possibility of exploiting configurational forces in the design of mechanical devices towards longitudinal actuation
from flexural vibrations of small amplitude at given frequency.
\end{abstract}


\section{Introduction}

   Configurational forces, introduced by Eshelby \cite{eshelby1, eshelby2, eshelby3} to describe the motion of defects in solids,
have been shown to act on elastic rods constrained by  sliding sleeves \cite{Bigoni2015MM}. In particular, the  sliding of an elastic rod through a frictionless sleeve
generates at both constraint ends an  \lq Eshelby-like' force of amount proportional  to the square  of the  bending moment   and direction parallel  to that of sliding.
Configurational forces may also be
derived through an asymptotic approach \cite{balabukh1970work, ballarini} or a material force balance \cite{hanna,oreilly,oreilly2,singh} and have been so far exploited in a series of novel applications \cite{Bigoni2014PRSA, Bigoni2014JMPS, bosiarmscale, bosidripping, bosiinjection, bosi2016asymptotic,  dal2017serpentine, liakou1,liakou2,liakou3}.
The action of Eshelby-like forces was recently disclosed in a dynamic framework and the case of a falling mass attached to an end of an elastic rod  investigated to reveal a complex and often counterintuitive motion  \cite{armanini2019}.

The dynamic characteristics of periodic structures is drawing an intensive research effort motivated by the design of metamaterials for wave guidance,  energy harvesting, vibration mitigation, shock absorbers, and earthquake protection \cite{baci2017, baci2018, bigonigei, brun, carta0, carta, carta2, casadei, deng, fraternali, garau, morinigei,
Nadkarni2014, Nadkarni2016, nievesIJSS, nievesJMPS,tallarico2017tilted, tallarico2017edge, tallarico2017propagation}. In  line with this research activity, the aim of the present article is to show how the nonlinearity introduced by configurational forces influences the dynamics of a periodic structural system. This periodic system is assumed to be made up of 
linear elastic beams of alternate stiffness, deforming both axially and flexurally, and constrained through different arrangements of sliding sleeves.
For periodic arrays of buckled beams (in the absence of configurational forces), non-linear wave propagation was addressed with the related localisation associated with  soliton-type solutions  \cite{Maurin2014_WM,Maurin2014_JSV,Maurin2016_WM1,Maurin2016_WM2}, but the effects of configurational forces was never considered. To this purpose, the conventional view on the Bloch waves as quasi-periodic solutions of linear spectral problems is replaced by the concept of `nested Bloch wave', which still maintains quasi-periodicity property but  is also consistent with the configurational forces at the ends of the periodically positioned sliding sleeves. As a result, the presence of a small flexural vibration is shown to overcome the band gap structure of the underlying problem of purely axial vibrations, so that longitudinal propagation may become possible at forbidden frequencies  when accompanied by transverse oscillation. Moreover, a small flexural oscillation is shown to induce axial resonance at certain frequencies, a situation in which the longitudinal displacements tend to become unbounded.
Therefore, configurational forces are shown in  this article to play a decisive role in the dynamics of a periodic structure. The exploitation of these concepts
may lead to the design of mechanical devices with band gap properties  tuned through transverse vibrations or
displaying a longitudinal actuation as the result of small amplitude flexural vibrations  at specific frequencies. 

The structure of the paper is as follows. The mechanical problem is introduced in Section \ref{sec:form} together with the new type of interface transmission conditions, which are nonlinear because they are related to the configurational forces at the sliding sleeves.  Quasi-periodic oscillations for the periodic structural system are investigated in Section \ref{sec:flex-sol} through a \lq nested Bloch-Floquet technique', so that compatible oscillations -- referred to as \lq nested Bloch waves' --  are obtained, connecting longitudinal and transverse waves. Finally, examples reported in Section \ref{sectresults} show how  axial vibrations may become possible in the presence of flexural oscillations and how the latter may induce resonance of the former.

\section{The dynamics of a periodic elastic structure with sliding sleeve constraints}\label{sec:form}

A periodic elastic system, subject to longitudinal (axial) and transverse (flexural) vibrations, is introduced in this section. The peculiarity of the system is the presence of sliding sleeve constraints, which generate concentrated axial reactions, the so-called \lq configurational' or \lq Eshelby'  forces, from flexural motion and introduce a nonlinearity in the dynamics.

\subsection{Geometry, mechanical properties and constraints}

\begin{figure}[h!]
\begin{center}
\includegraphics[width=\linewidth]{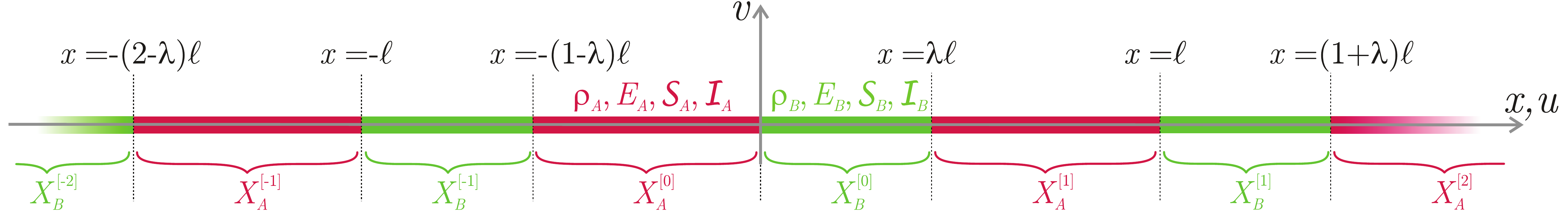}
	\caption{\label{fig:geom} A 1D structural system is obtained as the periodic repetition of the substructures $A$ (red) and $B$ (green), each of them  characterized by uniform properties of mass density $\rho$, Young modulus $E$, cross section area $\mathcal{S}$ and second moment of area $\mathcal{I}$. The two substructures have respective lengths $\ell_A=(1-\lambda)\ell$ and $\ell_B=\lambda\ell$. The dynamic behaviour of the system (in the spatial coordinate $x$ and time variable $t$) is analyzed in terms of the longitudinal  $u(x,t)$ and transverse $v(x,t)$ displacement components, so that both stretching  and bending are involved.}
\end{center}
\end{figure}

A one-dimensional structure, rectilinear in its undeformed state along the $x$-axis, is considered as the infinite repetition of a periodic piecewise constant distribution of mechanical properties,  Fig. \ref{fig:geom}. 
The repeated unit cell of length $\ell$ is a sequence of two substructures, $A$ and $B$, of respective length
\begin{equation}
\ell_A=(1-\lambda)\ell,\qquad
\ell_B=\lambda\ell,\qquad
\mbox{with}\,\,\lambda\in(0,1),
\end{equation} 
so that the continuous sets $X_A^{[m]}$ and $X_B^{[m]}$ are introduced as the collection of all the coordinates $x$ within the $m$-th unit cell ($m\in\mathbb{Z}$) associated with substructures $A$ and $B$, 
\begin{equation}\label{eq:sets0}
X^{[m]}_A=\left\{x~\big|\left(m-1+\lambda\right)\ell< x<m \ell \right\}~~~{\rm and}~~~
X^{[m]}_B=\left\{x~\big|m \ell < x<\left(m+\lambda\right)\ell\right\}, \qquad
m\in\mathbb{Z},
\end{equation} 
where,  for simplicity, the origin of the $x$-axis is assumed at the interface between the two substructures, so that the final coordinate of one substructure $A$  coincides with the initial coordinate of the following substructure $B$. It is  instrumental to introduce the discontinuous sets $X_A$ and $X_B$ as the collection of all the coordinates along the $x$-axis corresponding to the substructures $A$ and $B$, 
\begin{equation}
X_J=  \bigcup_{ m\in \mathbb{Z}} X_J^{[m]}, \qquad
J=A,B,
\end{equation}
and to define the coordinates of the interfaces between structures $A$ and $B$ with lower (-) and upper (+) bounds for the sets $X^{[m]}_{J}$, eqns \eqref{eq:sets0},
\begin{equation}
\partial^{-} X^{[m]}_{J}=\inf_{x} \left\{X^{[m]}_{J}\right\},
\qquad
\partial^{\text{\tiny{+}}} X^{[m]}_{J}=\sup_{x} \left\{X^{[m]}_{J}\right\},
\qquad
J=A,B,\,\,
 m\in \mathbb{Z},
\end{equation}
so that
\begin{equation}
\partial^{\text{\tiny{+}}} X^{[m]}_{A}=
\partial^{-} X^{[m]}_{B}= m \ell,
\qquad
\partial^{\text{\tiny{+}}} X^{[m]}_{B}=
\partial^{-} X^{[m+1]}_{A}=(m+\lambda) \ell,
\qquad
 m\in \mathbb{Z}.
\end{equation}

The inertial, geometrical, and mechanical properties are assumed to be uniform along each substructure, which is modeled as linearly elastic. Therefore, the distributions of (volumetric) mass density $\rho$, cross section area $\mathcal{S}$, second moment of area $\mathcal{I}$, and Young modulus $E$  are given by
\begin{equation}
\rho(x)=\rho_J\qquad
\mathcal{S}(x)=\mathcal{S}_J\qquad
\mathcal{I}(x)=\mathcal{I}_J\qquad
E(x)=E_J\qquad
\mbox{if}\,\,x\in X_J,\qquad
J=A,B.
\end{equation}

\begin{figure}[h!]
\begin{center}
\includegraphics[width=0.9\linewidth]{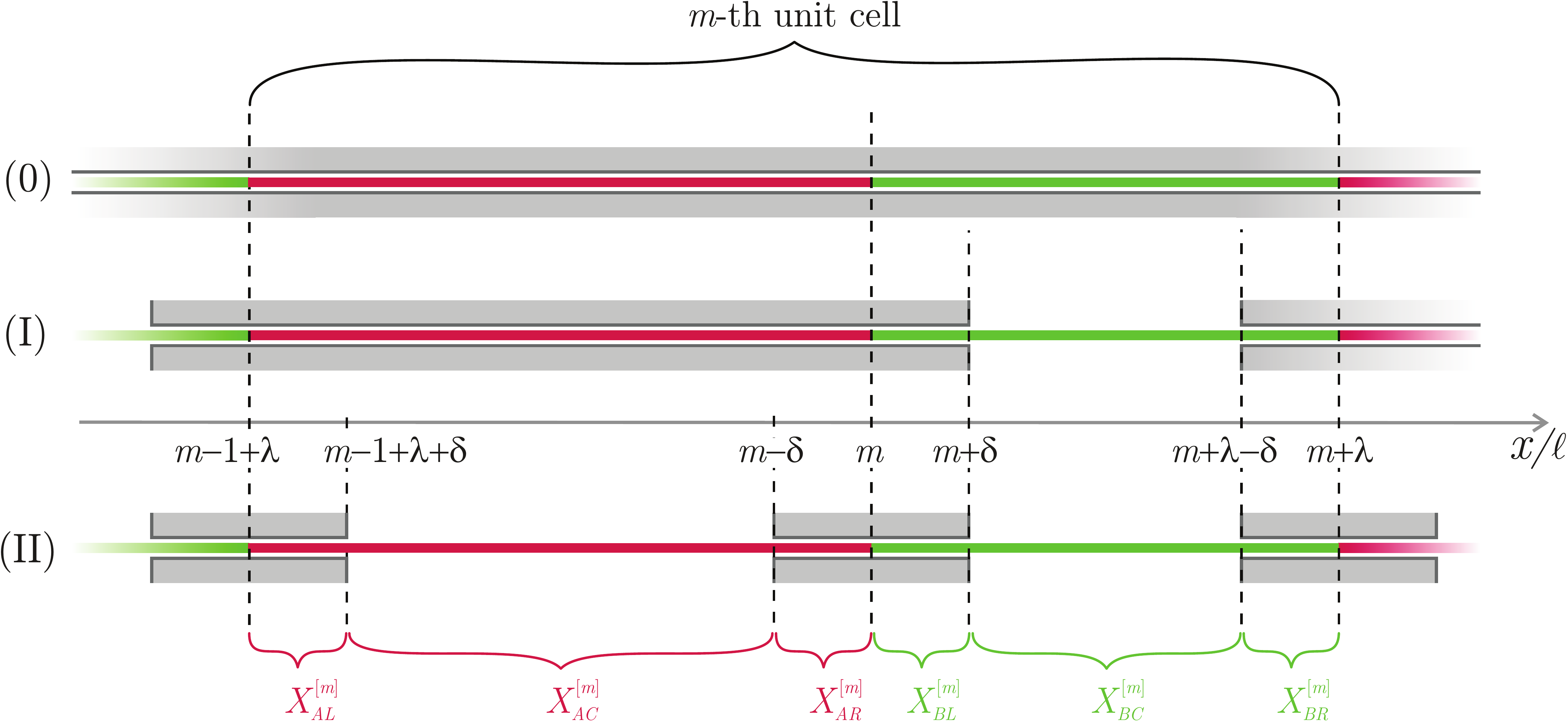}
\end{center}
	\caption{\label{fig:geom2} 
The three structural systems differ in the constraint applied to periodic structure shown in Fig. \ref{fig:geom}: System \textbf{(0)} only subject to longitudinal motion (constrained by an infinitely long sliding sleeve), 
System \textbf{(I)} where a sliding sleeve fully encloses the substructure $A$, but leaves transverse oscillations possible for the central part of the substructure $B$ (central part), 
and System \textbf{(II)} where a sliding sleeve symmetrically acts at the ends of the substructures, leaving to the central parts of both of them the possibility of flexural deformation (lower part).}
\end{figure}

A dynamic analysis is performed in the time variable $t$ for the periodic structural system, which undergoes planar motion, described under the assumption of linearized kinematics by the longitudinal $u(x,t)$ and transverse $v(x,t)$ displacement components, so that both stretching and flexure occur. 
In particular, three structural systems are considered, one of which is subject to purely longitudinal motion, while the other two involve the presence of sliding sleeve constraints, but of different extent, namely:
\begin{description}
\item[(0)] axially vibrating structure (corresponding to a sliding sleeve constraining  the whole structure, Fig. \ref{fig:geom2}, upper part)
\begin{equation}
v(x,t)=0,\qquad \forall \, x;
\end{equation}
\item[(I)] sliding sleeves of length $(1+2\delta-\lambda)\ell$ (with $\delta\in(0,\lambda/2)$) fully enclosing the substructure $A$  and  symmetrically  constraining the external parts of substructure $B$ (Fig. \ref{fig:geom2}, central part)
\begin{equation}
v(x,t)=0,\qquad x \in \left[\partial^{-} X^{[m]}_{A}-\delta \ell, \partial^{+} X^{[m]}_{A}+\delta \ell\right],\qquad
 m\in \mathbb{Z};
\end{equation}
\item[(II)] sliding sleeves of length $2\delta \ell$ (with $\delta\in(0,1/4)$), centered at the discontinuity points $\partial^{\pm} X^{[m]}_{J}$ (symmetrically constraining the external parts of substructures $A$ and $B$, Fig. \ref{fig:geom2}, lower part)
\begin{equation}
v(x,t)=0,\qquad x \in \left[\partial^{+} X^{[m]}_{J}-\delta \ell, \partial^{+} X^{[m]}_{J}+\delta \ell\right],\qquad
J=A,B,\,\,
 m\in \mathbb{Z}. 
\end{equation}
The presence of these constraints imply that the parameter $\lambda$ has to belong to the interval 
\begin{equation}\label{lambdarestricted}
\lambda\in(2\delta,1-2\delta).
\end{equation}
\end{description}

In the following, the analysis is developed for the structural System \textbf{(II)} and later exploited to investigate the behaviour of Systems \textbf{(0)} and \textbf{(I)} as particular cases. To this purpose, it is useful to  distinguish the three different subparts of the  substructure $J=A,B$ through the coordinate subsets $X^{[m]}_{JK}$, with  $K=C, L, R$ identifying the central, left, and right positions, as
\begin{equation}\label{eq:sets}
\resizebox{1\textwidth}{!}{$
\begin{array}{lll}
& \ds X^{[m]}_{AL}=\left\{x~\big|\left(m-1+\lambda\right)\ell< x<\left(m-1+\lambda+\delta\right) \ell \right\},\qquad
& \ds X^{[m]}_{BL}=\left\{x~\big|m \ell < x<\left(m+\delta\right)\ell\right\} ,
\\[4mm]
& X^{[m]}_{AC}=\left\{x~\big|\left(m-1+\lambda+\delta\right)\ell< x<(m-\delta) \ell \right\},\qquad
& X^{[m]}_{BC}=\left\{x~\big|(m+\delta) \ell < x<\left(m+\lambda-\delta\right)\ell\right\},\qquad
m\in\mathbb{Z},
\\[4mm]
& X^{[m]}_{AR}=\left\{x~\big|(m-\delta)\ell< x<m \ell \right\},\qquad
& X^{[m]}_{BR}=\left\{x~\big|\left(m+\lambda-\delta\right) \ell < x<\left(m+\lambda\right)\ell\right\},
\end{array}
$}
\end{equation}
so that their extremal coordinates respectively satisfy
\begin{equation}
\begin{array}{lll}
&\partial^{\text{\tiny{+}}} X^{[m]}_{AL}=
\partial^{-} X^{[m]}_{AC}= (m-1+\lambda+\delta)\ell,
\qquad\qquad
&\partial^{\text{\tiny{+}}} X^{[m]}_{AC}=
\partial^{-} X^{[m]}_{AR}= (m-\delta)  \ell,
\\[6mm]
&\partial^{\text{\tiny{+}}} X^{[m]}_{AR}=
\partial^{-} X^{[m]}_{BL}= m \ell,
\qquad\qquad
&\partial^{\text{\tiny{+}}} X^{[m]}_{BL}=
\partial^{-} X^{[m]}_{BC}= (m+\delta)\ell,
\qquad
\\[4mm]
&\partial^{\text{\tiny{+}}} X^{[m]}_{BC}=
\partial^{-} X^{[m]}_{BR}=  (m+\lambda-\delta)\ell,
\qquad\qquad
&\partial^{\text{\tiny{+}}} X^{[m]}_{BR}=
\partial^{-} X^{[m+1]}_{AL}= (m+\lambda) \ell,
\end{array}
\qquad
 m\in \mathbb{Z},
\end{equation}
and the discountinuous sets $X^{[m]}_{J}$ are given by
\begin{equation}
 X^{[m]}_{J}= \bigcup_{K=L,C,R} X^{[m]}_{JK},\qquad
 J=A,B,\qquad m \in \mathbb{Z}.
\end{equation}

\subsection{Equations of motion and boundary conditions\label{sec:nested_SW}}

Under the small displacement assumption, the free axial $u(x,t)$ and flexural $v(x,t)$ oscillations are governed by the well-known \textit{uncoupled} system of  partial differential equations with constant coefficients (where the rotational inertia is neglected) 
\begin{equation}\label{eq:gov-eq-cont}
\begin{cases}
\displaystyle c_J^2 u''(x,t)-\ddot u(x,t)=0,\\[4mm]
 \displaystyle c_J^2 {\cal R}^2_J v''''(x,t)+ \ddot v(x,t)=0,\\
\end{cases}  \qquad
x \in X_{JK},~~ J=A,B,~~ K=L,C,R,
\end{equation}
where a superscript ' or $\dot{~}$ denotes spatial or time derivative, 
${\cal R}_J=\sqrt{{\cal I}_J/{\cal S}_J}$  the radius of gyration and $c_J =\sqrt{E_J/\rho_J}$ the longitudinal wave speed  of the substructure $J$ $(J=A,B)$. 
With reference to System \textbf{(II)}, the boundary conditions are the following
\begin{itemize}
\item continuity of the longitudinal displacement field $u(x,t)$ at the subset boundaries
\begin{equation}\label{eq:BCs-sliding-fields0}
\resizebox{0.9\textwidth}{!}{$
\begin{array}{lll}
\ds\left.u(x,t)\right|_{x=\partial^{\text{\tiny{+}}} X^{[m]}_{AL}}=\left.u(x,t)\right|_{x=\partial^{-} X^{[m]}_{AC}},\qquad 
\ds\left.u(x,t)\right|_{x=\partial^{\text{\tiny{+}}} X^{[m]}_{BL}}=\left.u(x,t)\right|_{x=\partial^{-} X^{[m]}_{BC}},
\\[3mm]
\ds\left.u(x,t)\right|_{x=\partial^{\text{\tiny{+}}} X^{[m]}_{AC}}=\left.u(x,t)\right|_{x=\partial^{-} X^{[m]}_{AR}},
\quad 
\ds\left.u(x,t)\right|_{x=\partial^{\text{\tiny{+}}} X^{[m]}_{BC}}=\left.u(x,t)\right|_{x=\partial^{-} X^{[m]}_{BR}},
\\[3mm]
\ds\left.u(x,t)\right|_{x=\partial^{\text{\tiny{+}}} X^{[m]}_{AR}}=\left.u(x,t)\right|_{x=\partial^{-} X^{[m]}_{BL}},
\qquad
\ds\left.u(x,t)\right|_{x=\partial^{\text{\tiny{+}}} X^{[m]}_{BR}}=\left.u(x,t)\right|_{x=\partial^{-} X^{[m+1]}_{AL}},
\end{array}
\forall \,m \in \mathbb{Z};
$}
\end{equation}
\item continuity of the internal axial force $N(x,t)=E(x){\cal S}(x)
u'(x,t)$ at the interface between the two substructures
\begin{equation}\label{eq:BCs-sliding-fields00}
\begin{array}{lll}
 E_A{\cal S}_A\left.
\left[ u'(x,t)\right]\right|_{x=\partial^+ X_{AR}^{[m]}}
=
E_B{\cal S}_B \left.\left[u'(x,t)\right]\right|_{x=\partial^- X_{BL}^{[m]}},
\\[4mm]
\ds
E_B{\cal S}_B \left.\left[u'(x,t)\right]\right|_{x=\partial^+ X_{BR}^{[m]}}
=
 E_A{\cal S}_A\left.\left[ u'(x,t)\right]\right|_{x=\partial^- X_{AL}^{[m+1]}}
,
\end{array}
\qquad
\forall\, m\in \mathbb{Z};
\end{equation}
\item jump in the internal axial force $N(x,t)=E(x){\cal S}(x)
u'(x,t)$ at the interface between the two substructures given by the configurational force
\begin{equation}\label{eq:BCs-sliding-fields}
\begin{array}{lll}
\ds E_A{\cal S}_A\left.
\left[ u'(x,t)\right]\right|_{x=\partial^+ X_{AL}^{[m]}}
=
E_A{\cal S}_A \left.\left[u'(x,t)+\frac{{\cal R}_A^2 v''(x,t)^2}{2}\right]\right|_{x=\partial^- X_{AC}^{[m]}},
\\[4mm]
\ds
E_A{\cal S}_A \left.\left[u'(x,t)+\frac{{\cal R}_A^2 v''(x,t)^2}{2}\right]\right|_{x=\partial^+ X_{AC}^{[m]}}
=
 E_A{\cal S}_A\left.
\left[ u'(x,t)\right]\right|_{x=\partial^- X_{AR}^{[m]}},
\\[4mm]
\ds E_B{\cal S}_B\left.
\left[ u'(x,t)\right]\right|_{x=\partial^+ X_{BL}^{[m]}}
=
E_B{\cal S}_B \left.\left[u'(x,t)+\frac{{\cal R}_B^2 v''(x,t)^2}{2}\right]\right|_{x=\partial^- X_{BC}^{[m]}},
\\[4mm]
\ds
E_B{\cal S}_B \left.\left[u'(x,t)+\frac{{\cal R}_B^2 v''(x,t)^2}{2}\right]\right|_{x=\partial^+ X_{BC}^{[m]}}
=
 E_B{\cal S}_B\left.
\left[ u'(x,t)\right]\right|_{x=\partial^- X_{BR}^{[m]}},
\end{array}
\qquad
\forall\, m\in \mathbb{Z} .
\end{equation}
Note that the configurational force is given by the bending moment, $E_J{\cal S}_J {\cal R}_J^2 v''(x,t)$, evaluated at the sliding sleeve end, squared and divided by two times the beam's bending stiffness $E_J{\cal S}_J {\cal R}_J^2 $ ($J=A,B$);

\item null transverse displacement $v(x,t)$ and rotation $v'(x,t)$ within each sliding sleeve and at its ends
\begin{equation}\label{inc-inc}
\begin{array}{lll}
\ds v(x,t)=0,
\qquad
\ds v'(x,t)=0,
 \end{array}
\qquad
x\in X_{AL}^{[m]}\cup X_{AR}^{[m]}
\cup X_{BL}^{[m]}\cup X_{BR}^{[m]}\qquad m\in \mathbb{Z}.
\end{equation}
\end{itemize}

From the above boundary conditions is evident that:
\begin{itemize}
\item the transverse displacement, as well as its first derivative, is continuous along the periodic structure. The longitudinal displacement is continuous too, but its first derivative is discontinuous at each sliding sleeve end, due to the presence of the configurational force, and at each substructure's interface  where the material properties are discontinuous;
\item although the governing equations (\ref{eq:gov-eq-cont}) are uncoupled, a coupling  between the transverse $v(x,t)$ and the longitudinal $u(x,t)$ motion arises due to the boundary condition involving the configurational force. The coupling is weak because the former affects the latter, but not vice versa, and therefore longitudinal oscillations are \lq nested' into the transverse ones. Moreover, due to the nonlinear character of configurational forces, the presence of sliding sleeves introduces non-linear effects in structural  systems governed by linear differential equations.
\end{itemize}

\section{Nested Bloch-Floquet vibrations}
\label{sec:flex-sol}

Following standard techniques, the solution to the differential system (\ref{eq:gov-eq-cont}) can be sought through  separation of variables, so that both the displacement fields become functions of space, modulated by a function of time
\begin{equation}\label{separation}
u(x,t)  = U(x)\Phi(t),~~{\rm and}~~v(x,t)  = V(x)\Psi(t).
\end{equation}
Under this assumption, the system of two partial differential equations, eqn (\ref{eq:gov-eq-cont}), can be rewritten as two systems of ordinary differential equations, one in the time variable, the other in the space variable.  
The system in the time variable is given by
\begin{equation}\label{eq:gov-eq-cont20}
\begin{cases}
\displaystyle \ddot\Phi(t)+\omega^2\Phi(t)=0,\\[2mm]
\displaystyle \ddot\Psi(t)+\Omega^2\Psi(t)=0,
\end{cases}
\end{equation}
so that  the two functions $\Phi(t)$ and $\Psi(t)$ can be expressed in terms of sinusoidal and cosinusoidal trigonometric functions of $\omega t$ and $\Omega t$, with  $\omega$ and $\Omega$ being the angular frequencies for longitudinal and transverse vibrations, both for the moment unknown. Introducing these angular frequencies, the differential system in the space variable can be written as 
\begin{equation}\label{eq:gov-eq-cont2}
\begin{cases}
\displaystyle U''(x)+\frac{\omega^2}{c_J^2} U(x)=0,\\[2mm]
 \displaystyle V''''(x)-\frac{\Omega^2}{c_J^2 {\cal R}^2_J}  V(x)=0,
\end{cases}  \qquad
x \in X_{JK},~~ J=A,B,~~ K=L,C,R.
\end{equation}

Imposing the boundary conditions at all  upper and lower boundaries $\partial^{\text{\tiny{$\pm$}}} X^{[m]}_{JK}$ of each subset $X^{[m]}_{JK}$, 
system (\ref{eq:gov-eq-cont2}) leads to the following expressions for the spatial functions
\begin{equation}\label{eq:flex_sol}
\begin{array}{lll}
U(x)=\ds\sum_{m=-\infty}^{+\infty}
\sum_{J=A,B}
\sum_{K=L,C,R}
U_{JK}^{[m]}(x) \theta_{JK}^{[m]}(x) ,\qquad
V(x)=\ds\sum_{m=-\infty}^{+\infty}
\sum_{J=A,B}
V_{JC}^{[m]}(x) \theta_{JC}^{[m]}(x),
\end{array}
\end{equation}
where
\begin{equation}\label{eq:V1}
\resizebox{1\textwidth}{!}{$
\begin{array}{rll}
U_{JK}^{[m]}(x) &=&\ds\textsf{U}_{JK1}^{[m]} \sin\left(\frac{\omega}{c_J} 
\left(x-\partial^{-} X^{[m]}_{JK}\right)\right) + \textsf{U}_{JK2}^{[m]}\cos\left(\frac{\omega}{c_J}\left(x-\partial^{-} X^{[m]}_{JK}\right)\right),\\[4mm]
V_{JC}^{[m]}(x) &=&\textsf{V}_{JC1}^{[m]} \sin\left(\beta_J \left(x-\partial^{-} X^{[m]}_{JC}\right)\right) + \textsf{V}_{JC2}^{[m]}\cos\left(\beta_J\left(x-\partial^{-} X^{[m]}_{JC}\right)\right)\\[2mm]
&&
+\textsf{V}_{JC3}^{[m]}{\rm sinh}\left(\beta_J\left(x-\partial^{-} X^{[m]}_{JC}\right)\right)
+\textsf{V}_{JC4}^{[m]}{\rm cosh}\left(\beta_J \left(x-\partial^{-} X^{[m]}_{JC}\right))\right),\end{array}\qquad
\begin{array}{lll}
J=A,B,\\K=L,C,R,
\end{array}
$}
\end{equation}
with 
\begin{equation}\label{eq:beta_def}
\beta_J=\sqrt{\frac{\Omega}{c_J {\cal R}_J}},~~~J=A,B,
\end{equation}
and 
\begin{equation}
\theta_{JK}^{[m]}(x) =\theta\left(x-\partial^{-} X^{[m]}_{JC}\right) + \theta\left(\partial^{+} X^{[m]}_{JC}-x\right) -1,
\qquad J=A,B,\qquad K=L,C,R,\qquad m\in \mathbb{Z},
\end{equation}
in which $\theta(x)$ is the Heaviside step function, so that $\theta(x)=1$ for positive argument $x$ and $\theta(x)=0$ otherwise.

\subsection{Flexural vibrations}

Due to the weak coupling between axial and flexural vibrations, the latter vibrations can be solved independently of the former. More specifically, imposing the boundary conditions (\ref{inc-inc}) to the transverse displacement field (\ref{eq:flex_sol})$_2$ yields the following well-known eigenvalue problem for the parameter $\beta_J$
\begin{equation}\label{eq:solv_cond}
\cosh\left(\beta_J(\ell_J-2\delta\ell)\right) \cos\left(\beta_J(\ell_J-2\delta\ell)\right)=1, ~~~ J=A,B ,
\end{equation}
providing an infinite set of solutions 
\begin{equation}\label{betavalues}
\beta_J(n_J)=\dfrac{\Xi^{(n_J)}}{(\ell_J-2\delta\ell)}\qquad \mbox{with}\qquad \Xi^{(n_J)}\approx\left\{
\begin{array}{lll}
4.730,\qquad &n_J=1,\\
\left(n_J+\frac{1}{2}\right)\pi,\qquad &n_J\geq 2,
\end{array}
\right.\qquad
J=A,B,
\end{equation}
each corresponding to the following eigenvector 
\begin{equation}\label{eq:def_Dn}
\textsf{V}_{JC2}^{[m]}=
-\textsf{V}_{JC4}^{[m]}=
-\frac{{\rm sin
	}(\beta_J (\ell_J-2\delta\ell))-{\rm sinh}(\beta_J (\ell_J-2\delta\ell))}{\cos{(\beta_J (\ell_J-2\delta\ell))}-{\cosh(\beta_J (\ell_J-2\delta\ell))}}\textsf{V}_{JC1}^{[m]},
\qquad
\textsf{V}_{JC3}^{[m]}=
-\textsf{V}_{JC1}^{[m]}~~~~ J=A,B.
\end{equation}

It is important to observe that the present analysis leads to the independent evaluation  of the  eigenvalues $\beta_A(n_A)$ and $\beta_B(n_B)$. However, for System \textbf{(II)}, due to the axial interaction between the substructures, these two eigenvalues are related to the same angular frequency $\Omega$, eqn (\ref{eq:beta_def}), so that the free oscillation may be realized  only when 
\begin{equation}
\beta_A^2 c_A {\cal R}_A=
\beta_B^2 c_B {\cal R}_B,
\end{equation}
a condition imposing that a transverse motion governed by modes $n_A$ and $n_B$ becomes possible for  two substructures  only when 
\begin{equation}\label{vinculo}
\frac{c_A {\cal R}_A}{c_B {\cal R}_B} \left(\frac{\lambda-2\delta}{1-\lambda-2\delta}\right)^2=
\left(\frac{\Xi^{(n_B)}}{\Xi^{(n_A)}}\right)^2,
\end{equation}
implying the following linear relation\footnote{
Equation (\ref{vinculo}) has two solutions: one is given by eqn (\ref{length_constraint}) and always satisfies the geometrical restriction expressed by eqn (\ref{lambdarestricted}), while the other solution violates the latter restriction, so that it is mechanically meaningless.
} 
between $\lambda$ and $\delta$
\begin{equation}\label{length_constraint}
\lambda=2 \delta+\frac{1-4\delta}
{1+\frac{\Xi^{(n_A)}}{\Xi^{(n_B)}}\sqrt{\frac{c_A {\cal R}_A}{c_B {\cal R}_B}}}\in(2\delta,1-2\delta).
 \end{equation}

\begin{figure}[t!]
\begin{center}
\includegraphics[width=0.5\linewidth]{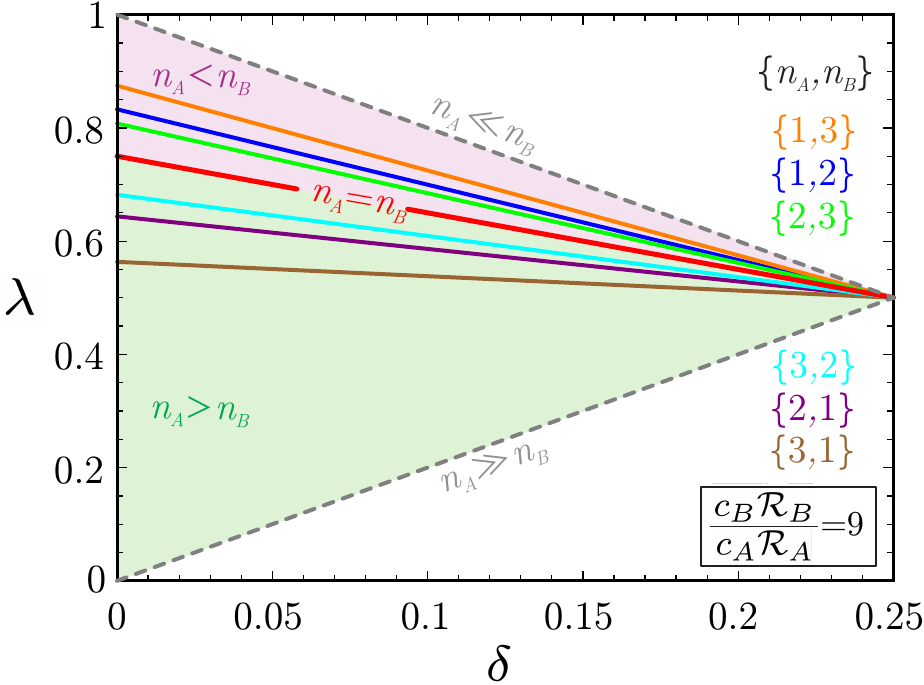}
\end{center}
\caption{\label{lambda_delta_r} 
Parameter $\lambda$ as a function of the dimensionless sliding sleeve half-length $\delta$, eqn (\ref{length_constraint}), allowing for transverse oscillations  in both substructures of System (\textbf{II}) under modes $n_A$ and $n_B$. The linear relation has been drawn only for the  mode pairs $\{n_A,n_B\}$ with $n_A$ and $n_B$ ranging between 1 and 3 of a structural system with $(c_B \mathcal{R}_B)/(c_A \mathcal{R}_A)=9$. In particular, the red line represents the case $n_A=n_B=n$, under which the two substructures may simultaneously oscillate under every mode $n$. This line and the two dashed lines, the latters representing upper and lower bounds for $\lambda$ (respectively given  by the limit cases $n_A\ll n_B$ and $n_A\gg n_B$), define two regions in the plane where all the cases $n_A<n_B$ (light red background) and $n_A>n_B$ (light green background) lie. The specific mode pairs $\{n_A,n_B\}$, to which the other colored straight lines are referred, are listed in the figure with corresponding colors.}
\end{figure}

The geometrical constraint (\ref{length_constraint}) shows that, for finite values of $\sqrt{(c_B \mathcal{R}_B)/(c_A \mathcal{R}_A)}$, the lower (upper) bound for $\lambda$ is attained when $n_A \gg n_B$ ($n_A \ll n_B$).
The linear relation between $\lambda$ and $\delta$ defined by equation 
 (\ref{length_constraint}) is graphically represented in Fig.  \ref{lambda_delta_r}, 
for $(c_B \mathcal{R}_B)/(c_A \mathcal{R}_A)=9$ and different   mode pairs $\{n_A,n_B\}$ (with $n_A$ and $n_B$ ranging between 1 and 3).

It is remarked that the  relation (\ref{length_constraint}) holds only for System \textbf{(II)} when both substructures are vibrating transversally. 
Whenever at least one of the two substructures does not display transverse oscillations (as it occurs for System \textbf{(II)} when $\textsf{V}_{AC1}^{[m]}=0$ or $\textsf{V}_{BC1}^{[m]}=0$ for every $m$, but also for Systems \textbf{(0)} and \textbf{(I)}), relation (\ref{length_constraint}) does not hold any longer.

With reference again to System (\textbf{II}), it is worth to note that when $\lambda$ is such that the two substructures may simultaneously oscillate under transverse modes, respectively, $n_A$ and $n_B$:
\begin{itemize}
\item if $n_A\neq n_B$ then this is the unique pair of modes for which transverse oscillation may occur;
\item if $n_A= n_B=n$ then the two substructures may   oscillate transversally for every mode $n$. In this case, the periodic structure has a length parameter $\lambda$ given as a function of the sliding sleeve half-length $\delta$,  the longitudinal wave speeds $c_J$  and the radii of gyration ${\cal R}_J$ of the two substructures ($J=A,B$) in the form
\begin{equation}
\lambda=2 \delta+\frac{1-4\delta}
{1+\sqrt{\frac{c_A {\cal R}_A}{c_B {\cal R}_B}}}.
\end{equation}
\end{itemize}

\subsection{Axial vibrations}

In order to solve for axial oscillations, the boundary conditions (\ref{eq:BCs-sliding-fields0}), (\ref{eq:BCs-sliding-fields00}), and (\ref{eq:BCs-sliding-fields}) involving the longitudinal displacement may be rewritten via separation of variables, eqn (\ref{separation}), and using the expression (\ref{eq:flex_sol}) for the spatial functions. In particular, the continuity of the longitudinal displacement (\ref{eq:BCs-sliding-fields0}) implies
\begin{equation}\label{reducedcont}
\resizebox{1\textwidth}{!}{$
\begin{array}{lll}
\ds \textsf{U}_{AC2}^{[m]}=\textsf{U}_{AL1}^{[m]} \sin\left(\frac{\omega \delta \ell}{c_A} \right) + \textsf{U}_{AL2}^{[m]}\cos\left(\frac{\omega\delta \ell}{c_A}\right),\qquad 
\ds \textsf{U}_{AR2}^{[m]}=\textsf{U}_{AC1}^{[m]} \sin\left(\frac{\omega (1-\lambda-2\delta) \ell}{c_A} \right) + \textsf{U}_{AC2}^{[m]}\cos\left(\frac{\omega(1-\lambda-2\delta)  \ell}{c_A}\right),\\[4mm]
\ds \textsf{U}_{BL2}^{[m]}=\textsf{U}_{AC1}^{[m]} \sin\left(\frac{\omega \delta \ell}{c_A} \right) + \textsf{U}_{AC2}^{[m]}\cos\left(\frac{\omega\delta \ell}{c_A}\right),\qquad
\ds \textsf{U}_{BC2}^{[m]}=\textsf{U}_{BL1}^{[m]} \sin\left(\frac{\omega \delta \ell}{c_B} \right) + \textsf{U}_{BL2}^{[m]}\cos\left(\frac{\omega\delta \ell}{c_B}\right),\\[4mm]
\ds \textsf{U}_{BR2}^{[m]}=\textsf{U}_{BC1}^{[m]} \sin\left(\frac{\omega (\lambda-2\delta) \ell}{c_B} \right) + \textsf{U}_{BC2}^{[m]}\cos\left(\frac{\omega(\lambda-2\delta)  \ell}{c_B}\right),\qquad
\ds \textsf{U}_{AL2}^{[m+1]}=\textsf{U}_{BR1}^{[m]} \sin\left(\frac{\omega \delta \ell}{c_B} \right) + \textsf{U}_{BR2}^{[m]}\cos\left(\frac{\omega\delta \ell}{c_B}\right),
\end{array}
$}
\end{equation}
while the continuity equations (\ref{eq:BCs-sliding-fields00}) implies
 \begin{equation}\label{0003}\begin{array}{ll}
\ds E_A\mathcal{S}_A\frac{\omega}{c_A}\left[
\textsf{U}_{AR1}^{[m]} \cos\left(\frac{\omega\delta\ell}{c_A} \right) - \textsf{U}_{AR2}^{[m]}\sin\left(\frac{\omega \delta\ell}{c_A}\right)
\right]=
\ds E_B\mathcal{S}_B\frac{\omega}{c_B}
\textsf{U}_{BL1}^{[m]} ,
\\ [4mm]
\ds E_B\mathcal{S}_B\frac{\omega}{c_B}\left[
\textsf{U}_{BR1}^{[m]} \cos\left(\frac{\omega\delta\ell}{c_B} \right) - \textsf{U}_{BR2}^{[m]}\sin\left(\frac{\omega \delta\ell}{c_B}\right)
\right]=
\ds E_A\mathcal{S}_A\frac{\omega}{c_A}
\textsf{U}_{AL1}^{[m+1]} .
\end{array}
\end{equation}   
Moreover, holding the jump conditions (\ref{eq:BCs-sliding-fields}) at each sliding sleeve end $\partial X^{(\pm)}_{J}$ ($J=A,B$) for every time $t$, the frequency of the longitudinal oscillation is related to that of the transverse through
\begin{equation}\label{eq:long-freq}
\omega = 2 \Omega,
\end{equation}   
a condition defining that the two types of oscillation are \lq nested'. It follows that the jump conditions (\ref{eq:BCs-sliding-fields})  at each sliding sleeve end  $\partial X^{(\pm)}_{J}$ ($J=A,B$)  reduce to
 \begin{equation}\label{0001}
\resizebox{1\textwidth}{!}{$
 \begin{array}{ll}
 \begin{array}{ll}
\ds \frac{\omega}{c_A}\left[
\textsf{U}_{AL1}^{[m]} \cos\left(\frac{\omega\delta\ell}{c_A} \right) - \textsf{U}_{AL2}^{[m]}\sin\left(\frac{\omega\delta\ell}{c_A}\right)
\right]=
\ds \frac{\omega}{c_A}
\textsf{U}_{AC1}^{[m]} 
+
\frac{2 \Omega^2}{c_A^2} \left(\textsf{V}_{AC1}^{[m]}\right)^2 \left[
\frac{\sin[(1-\lambda-2\delta) \beta_A\ell]-\sinh[(1-\lambda-2\delta) \beta_A\ell]}{\cos[(1-\lambda-2\delta)\beta_A\ell]-\cosh[(1-\lambda-2\delta)\beta_A \ell]}\right]^2,
\end{array}
\\[8mm]
\begin{array}{ll}
\ds \frac{\omega}{c_A}\left[
\textsf{U}_{AC1}^{[m]} \cos\left(\frac{\omega(1-\lambda-2\delta)\ell}{c_A} \right) - \textsf{U}_{AC2}^{[m]}\sin\left(\frac{\omega(1-\lambda-2\delta)\ell}{c_A}\right)
\right]\\[4mm]
\ds+
\frac{2\Omega^2}{c_A^2} \left(\textsf{V}_{AC1}^{[m]}\right)^2 \left[
\frac{\sin[(1-\lambda-2\delta) \beta_A\ell] \cosh[(1-\lambda-2\delta)\beta_A \ell]
-\sinh[(1-\lambda-2\delta) \beta_A\ell] \cos[(1-\lambda-2\delta)\beta_A\ell]}{\cos[(1-\lambda-2\delta)\beta_A\ell]-\cosh[(1-\lambda-2\delta)\beta_A \ell]}\right]^2
=
\ds \frac{\omega}{c_A}
\textsf{U}_{AR1}^{[m]},
\end{array}
\\[12mm]
\begin{array}{ll}
\ds \frac{\omega}{c_B}\left[
\textsf{U}_{BL1}^{[m]} \cos\left(\frac{\omega\delta\ell}{c_B} \right) - \textsf{U}_{BL2}^{[m]}\sin\left(\frac{\omega\delta\ell}{c_B}\right)
\right]=
\ds \frac{\omega}{c_B}
\textsf{U}_{BC1}^{[m]} 
+
\frac{2 \Omega^2}{c_B^2} \left(\textsf{V}_{BC1}^{[m]}\right)^2 \left[
\frac{\sin[(\lambda-2\delta) \beta_B\ell]-\sinh[(\lambda-2\delta) \beta_B\ell]}{\cos[(\lambda-2\delta)\beta_B\ell]-\cosh[(\lambda-2\delta)\beta_B \ell]}\right]^2,
\end{array}
\\[8mm]
\begin{array}{ll}
\ds \frac{\omega}{c_B}\left[
\textsf{U}_{BC1}^{[m]} \cos\left(\frac{\omega(\lambda-2\delta)\ell}{c_B} \right) - \textsf{U}_{BC2}^{[m]}\sin\left(\frac{\omega(\lambda-2\delta)\ell}{c_B}\right)
\right]\\[4mm]
\ds+
\frac{2\Omega^2}{c_B^2} \left(\textsf{V}_{BC1}^{[m]}\right)^2 \left[
\frac{\sin[(\lambda-2\delta) \beta_B\ell] \cosh[(\lambda-2\delta)\beta_B \ell]
-\sinh[(\lambda-2\delta) \beta_B\ell] \cos[(\lambda-2\delta)\beta_B\ell]}{\cos[(\lambda-2\delta)\beta_B\ell]-\cosh[(\lambda-2\delta)\beta_B \ell]}\right]^2
=
\ds \frac{\omega}{c_B}
\textsf{U}_{BR1}^{[m]}. 
\end{array}
\end{array}
$}
\end{equation}

\subsection{Bloch-Floquet analysis and resonance}

Free oscillations are  sought with quasi-periodicity properties by imposing the Bloch-Floquet condition with phase angle $\phi$ between two subsequent structural unit cells
\begin{equation}\label{quasi-periodicity}
u(x+p \ell,t)=e^{i\, p  \phi} \, u(x,t),\qquad
v(x+p \ell,t)=e^{i\, p  \phi/2} \,  v(x,t),\qquad 
p\in \mathbb{Z},
\end{equation}
where $i$ is the imaginary unit.
Considering the separation of variables, eqn (\ref{separation}), the quasi-periodicity provides the following constraints on the (complex) amplitudes defining the motion in the different cells
\begin{equation}
\begin{array}{lll}
\textsf{U}_{JKj}^{[p]}=e^{i\, p  \phi} \, \textsf{U}_{Jj}^{[0]},\\
\textsf{V}_{JCk}^{[p]}=e^{i\, p  \phi/2} \, \textsf{V}_{Jk}^{[0]},
\end{array}
\qquad 
p\in \mathbb{Z},\,\,\,
J=A,B,\,\,\,K=L,C,R,\,\,\,
j=1,2,\,\,\,k=1,2,3,4,
\end{equation}
so that the quasi-periodic motion of the infinite periodic structural system is defined once the amplitudes defining the motion of the $0$-th cell are given. In  particular, the application of the Bloch-Floquet condition to the boundary conditions (\ref{reducedcont}), (\ref{0003}), and (\ref{0001}) leads to the following  linear system in the twelve amplitudes of the axial motion in the $0$-th cell, $\textsf{U}_{JKj}^{[0]}$ ($J=A,B$, $K=L,C,R,$, and $j=1,2$), to be solved for given values of the two transverse amplitudes $V_{AC1}^{[0]}$ and $V_{BC1}^{[0]}$, 
both corresponding respectively to modes $n_A$ and $n_B$,
\begin{equation}\label{eq:kernel00}
\mathbb{M}
\frac{\textbf{\textsf{U}}^{[0]}}{\ell}
=
\frac{1}{2}\frac{\omega \ell}{c_A}
\left\{
\boldsymbol{\Gamma}_A
\left(\frac{\textsf{\textsf{V}}_{AC1}^{[0]}}{\ell}\right)^2
+\frac{c_A}{c_B}
\boldsymbol{\Gamma}_B
\left(\frac{\textsf{V}_{BC1}^{[0]}}{\ell}\right)^2
\right\},
\end{equation} 
which,  under the quasi-periodicity assumption (\ref{quasi-periodicity}),  also includes the boundary conditions (\ref{reducedcont}), (\ref{0003}), and (\ref{0001}) for every cell $p\in\mathbb{Z}$.
In eqn (\ref{eq:kernel00}),  $\mathbb{M}$ is a 12 by 12 matrix, whose non-null coefficients are
\begin{equation}
\begin{array}{ccc}
\ds \mathbb{M}_{1,1}=\mathbb{M}_{3,5}=-\mathbb{M}_{7,2}=-\mathbb{M}_{9,6}=\sin\left(\frac{\omega \delta \ell}{c_A}\right),\qquad 
\ds \mathbb{M}_{1,2}=\mathbb{M}_{3,6}=\mathbb{M}_{7,1}= \mathbb{M}_{9,5}=  \cos\left(\frac{\omega \delta \ell}{c_A}\right),\\[4mm]
\ds \mathbb{M}_{1,4}=\mathbb{M}_{2,6}=\mathbb{M}_{3,8}=\mathbb{M}_{4,10}=\mathbb{M}_{5,12}=\mathbb{M}_{7,3}=\mathbb{M}_{8,5}=\mathbb{M}_{10,9}=\mathbb{M}_{11,11}=-1,
\\[4mm]
\ds \mathbb{M}_{4,7}=\mathbb{M}_{6,11}=-\mathbb{M}_{10,8}=-\mathbb{M}_{12,12}=\sin\left(\frac{\omega \delta \ell}{c_B}\right),\qquad 
\mathbb{M}_{4,8}=\mathbb{M}_{6,12}=\mathbb{M}_{10,7}=\mathbb{M}_{12,11}= \cos\left(\frac{\omega \delta \ell}{c_B}\right),\\[4mm]
\ds \mathbb{M}_{2,3}=-\mathbb{M}_{8,4}=\sin\left(\frac{\omega(1-\lambda -2\delta)\ell}{c_A}\right),
\qquad
\ds \mathbb{M}_{2,4}=\mathbb{M}_{8,3}=\cos\left(\frac{\omega(1-\lambda -2\delta)\ell}{c_A}\right),\\[4mm]
\ds \mathbb{M}_{5,9}=-\mathbb{M}_{11,10}=\sin\left(\frac{\omega(\lambda -2\delta)\ell}{c_B}\right),
\qquad
\ds \mathbb{M}_{5,10}=\mathbb{M}_{11,9}=\cos\left(\frac{\omega(\lambda -2\delta)\ell}{c_B}\right),
\\[4mm]
\ds \mathbb{M}_{12,1}=-\frac{E_A \mathcal{S}_A}{E_B \mathcal{S}_B}\frac{c_B}{c_A} e^{i\,  \phi} \, ,
\qquad
\mathbb{M}_{6,2}=- e^{i\,  \phi} \, ,
\qquad
\mathbb{M}_{9,7}=-\frac{E_B \mathcal{S}_B}{E_A \mathcal{S}_A}\frac{c_A}{c_B},
\end{array}
\end{equation}
while the three vectors $\textbf{\textsf{U}}^{[0]}$, $\boldsymbol{\Gamma}_A$, and $\boldsymbol{\Gamma}_B$ are defined as
\begin{equation}
\begin{array}{lll}
\ds\textbf{\textsf{U}}^{[0]}=\left[\textsf{U}_{AL1}^{[0]},
\textsf{U}_{AL2}^{[0]},
\textsf{U}_{AC1}^{[0]}, 
\textsf{U}_{AC2}^{[0]}, 
\textsf{U}_{AR1}^{[0]}, 
\textsf{U}_{AR2}^{[0]},
\textsf{U}_{BL1}^{[0]}, 
\textsf{U}_{BL2}^{[0]}, 
\textsf{U}_{BC1}^{[0]}, 
\textsf{U}_{BC2}^{[0]}, \textsf{U}_{BR1}^{[0]}, \textsf{U}_{BR2}^{[0]}\right]^T,
\\[2mm]
\ds\boldsymbol{\Gamma}_A=\left[0,0, 0, 0, 0, 0, \Gamma_{A1}, \Gamma_{A2}, 0, 0, 0,0\right]^T,\\[2mm]
\ds\boldsymbol{\Gamma}_B=\left[0,0, 0, 0, 0, 0, 0, 0, 0, \Gamma_{B1}, \Gamma_{B2},0\right]^T,
\end{array}
\end{equation}
where the superscript $T$ denotes the transpose operator and the coefficients $\Gamma_{Jk}$ ($J=A,B$ and $k=1,2$) are given by
\begin{equation}\label{final_system}
\begin{array}{lll}
\ds \Gamma_{J1}=\left[
\frac{\sin[\Xi^{(n_J)}]-\sinh[\Xi^{(n_J)}]}
{\cos[\Xi^{(n_J)}]-\cosh[\Xi^{(n_J)}]}\right]^2,\\[4mm]
\ds \Gamma_{J2}=-\left[
\frac{\sin[\Xi^{(n_J)}] \cosh[(\Xi^{(n_J)}]
-\sinh[\Xi^{(n_J)}] \cos[\Xi^{(n_J)}]}
{\cos[\Xi^{(n_J)}]-\cosh[\Xi^{(n_J)}]}\right]^2,
\end{array}
\qquad
J=A,B.
\end{equation}
From the linear equation (\ref{eq:kernel00}) it is evident that the analysis developed for System  \textbf{(II)} also includes, as particular cases, that for  System \textbf{(0)} (by considering  $\textsf{V}_{AC1}^{[0]}=\textsf{V}_{BC1}^{[0]}=0$) and System \textbf{(I)} (by considering  $\textsf{V}_{AC1}^{[0]}=0$).

For the given transverse amplitudes  $\textsf{V}_{AC1}^{[0]}$ and $\textsf{V}_{BC1}^{[0]}$, the longitudinal motion of the system can be solved 
when the matrix $\mathbb{M}$ in Eq. \eqref{eq:kernel00} is not singular, in other words, when 
\begin{equation}\label{determinante12x12}
\resizebox{1\textwidth}{!}{$
\begin{array}{lll}
\det[\mathbb{M}]=e^{i\,\phi} \ds\left[2\cos\left(\frac{\omega(1-\lambda)\ell}{c_A}\right)\cos\left(\frac{\omega\lambda\ell}{c_B}\right)-\left(
\frac{c_A}{c_B}\frac{E_B S_B}{E_A S_A}
+\frac{c_B}{c_A}\frac{E_A S_A}{E_B S_B}
\right)
\sin\left(\frac{\omega(1-\lambda)\ell}{c_A}\right)\sin\left(\frac{\omega\lambda\ell}{c_B}\right)
- 2\cos \phi\right],
\end{array}
$}
\end{equation}
is not null, a condition 
showing that the sliding sleeve length parameter $\delta$ is not involved. It follows that the vanishing of the determinant is attained when the pair $\omega$ and $\phi$ satisfies\footnote{The annihilation condition (\ref{annihilation_determinante12x12}) is equivalent to the well-known dispersion equation for one-dimensional scalar (out-of plane) Bloch-Floquet wave in a periodic bi-material reported in \cite{movchan2002asymptotic} (Chapt. 1,  equation (1.170)), where the shear moduli of the bi-material are replaced by the respective Young moduli.}
\begin{equation}\label{annihilation_determinante12x12}
\resizebox{1\textwidth}{!}{$
\ds2\cos\left(\frac{\omega(1-\lambda)\ell}{c_A}\right)\cos\left(\frac{\omega\lambda\ell}{c_B}\right)-\left(
\frac{c_A}{c_B}\frac{E_B S_B}{E_A S_A}
+\frac{c_B}{c_A}\frac{E_A S_A}{E_B S_B}
\right)
\sin\left(\frac{\omega(1-\lambda)\ell}{c_A}\right)\sin\left(\frac{\omega\lambda\ell}{c_B}\right)
= 2\cos \phi,
$}
\end{equation}
a relation which mathematically defines the singularity of the matrix $\mathbb{M}$ and from a mechanical point of view represents the  \textit{resonance condition} of the  periodic structural system. For such a condition an infinitely large longitudinal motion  results from an infinitely small transverse oscillations.

\section{Results on the dynamics of a periodic structure subject to configurational forces}
\label{sectresults}
The  theoretical framework developed in the previous sections is  exploited here to investigate the dynamic behaviour of the considered three periodic structures subject to configurational forces: Systems \textbf{(0)}, \textbf{(I)}, and \textbf{(II)}.  

\subsection{Purely axial vibrations: System \textbf{(0)}}\label{type0}

When transverse motion is excluded, $v(x,t)=0$, only longitudinal oscillations become possible. A way to eliminate flexural vibrations is to insert the whole structure inside a sliding sleeve, so that now the definition of the subsets $X_{JL}$, $X_{JC}$, and $X_{JR}$ ($J=A,B$) becomes meaningless and the longitudinal displacement field can be expressed as
\begin{equation}\label{eq:flex_solapp}
\begin{array}{lll}
U(x)=\ds\sum_{m=-\infty}^{+\infty}
\sum_{J=A,B}
U_{J}^{[m]}(x) \theta_{J}^{[m]}(x) ,
\end{array}
\end{equation}
where
\begin{equation}\label{eq:V1app}
\begin{array}{rll}
U_{J}^{[m]}(x) &=&\ds\textsf{U}_{J1}^{[m]} \sin\left(\frac{\omega}{c_J} 
\left(x-\partial^{-} X^{[m]}_{J}\right)\right) + \textsf{U}_{J2}^{[m]}\cos\left(\frac{\omega}{c_J}\left(x-\partial^{-} X^{[m]}_{J}\right)\right),\end{array}
\begin{array}{lll}
J=A,B,\\
\end{array}
\end{equation}
with 
\begin{equation}
\theta_{J}^{[m]}(x) =\theta\left(x-\partial^{-} X^{[m]}_{J}\right) + \theta\left(\partial^{\text{\tiny{+}}} X^{[m]}_{J}-x\right) -1,
\qquad J=A,B,\qquad m\in \mathbb{Z}.
\end{equation}

Due to the simplicity of expression (\ref{eq:flex_solapp}) for the longitudinal field, the twelve boundary conditions (\ref{eq:BCs-sliding-fields0}), (\ref{eq:BCs-sliding-fields00}), and (\ref{eq:BCs-sliding-fields}) reduce to four continuity conditions at the interfaces between the two substructures, which considering the quasi-periodicity condition (\ref{quasi-periodicity}), are equivalent to the following homogeneous  linear system for the four amplitudes $\textsf{U}_{A1}^{[0]}$, $\textsf{U}_{A2}^{[0]}$, $\textsf{U}_{B1}^{[0]}$, and $\textsf{U}_{B2}^{[0]}$
\begin{equation}\label{eq:kernel2a}
\begin{pmatrix}
\ds \sin\left( \frac{\omega (1-\lambda)\ell}{c_A}\right) 
&\ds \cos\left(  \frac{\omega (1-\lambda)\ell}{c_A} \right)
&0
&-1\\[1mm]
\ds 0
&\ds - e^{i\,  \phi}
&\ds \sin\left( \frac{\omega \lambda\ell}{c_B}\right) 
&\ds\cos\left(  \frac{\omega \lambda\ell}{c_B} \right)
\\[1mm]
\ds
\cos\left(\frac{\omega(1-\lambda) \ell}{c_A} \right)
&\ds -\sin\left( \frac{\omega(1-\lambda) \ell}{c_A} \right) 
&\ds -\frac{E_B \mathcal{S}_B}{E_A \mathcal{S}_A}
\frac{c_A} {c_B}
&0\\[1mm]
\ds
-\frac{E_A \mathcal{S}_A}{E_B \mathcal{S}_B}
\frac{c_B}{c_A} e^{i\,  \phi}
&0
&\ds \cos\left(\frac{\omega \lambda \ell}{c_B} \right)
&\ds -\sin\left(\frac{\omega \lambda \ell}{c_B} \right) 
\end{pmatrix}
\left[
\begin{array}{ccc}
\textsf{U}_{A1}^{[0]}\\[4mm]
\textsf{U}_{A2}^{[0]}\\[4mm]
\textsf{U}_{B1}^{[0]}\\[4mm]
\textsf{U}_{B2}^{[0]}
\end{array}
\right]
=
\left[
\begin{array}{ccc}
0\\
0\\
0\\
0
\end{array}
\right].
\end{equation} 
It is noted that the determinant of the matrix involved in the above linear problem coincides with that of the matrix $\mathbb{M}$, eqn (\ref{determinante12x12}), so that non-trivial solutions (namely, longitudinal oscillations) exist when eqn (\ref{annihilation_determinante12x12}) holds, defining the dispersion diagram  $\omega$ - $\phi$ at varying the properties of the substructures.

For the sake of simplicity, the results are presented for this and the other two systems assuming the same mass density and geometrical properties for the two substructures
\begin{equation}\label{eq:simple-par}
\rho_J=\overline{\rho}\qquad
\mathcal{S}_J=\overline{\mathcal{S}}\qquad
\mathcal{I}_J=\overline{\mathcal{I}}\qquad
\mbox{if}\,\,x\in X_J,\qquad
J=A,B,
\end{equation}
but different stiffness defined  in relation to a positive parameter $r$
\begin{equation}\label{eq:par1-I}
E_A=\overline{E}\qquad
E_B=r^2\overline{E},
\end{equation}
so that the longitudinal wave speed for each substructure is given by
\begin{equation}\label{eq:par2-I}
c_A=\overline{c}\qquad
c_B=r \overline{c},\qquad
\end{equation}
with $\overline{c}=\sqrt{\overline{E}/\overline{\rho}}$. 
Under this assumption, the matrix determinant (\ref{annihilation_determinante12x12}) reduces to
\begin{equation}\label{detridotto}
2\cos\left(\frac{\omega(1-\lambda)\ell}{\overline{c}}\right)\cos\left(\frac{\omega\lambda\ell}{r\overline{c}}\right)-\left(
r+\frac{1}{r}\right)
\sin\left(\frac{\omega(1-\lambda)\ell}{\overline{c}}\right)\sin\left(\frac{\omega\lambda\ell}{r \overline{c}}\right)
=2 \cos \phi,
\end{equation}
which defines the relation between the Bloch-wave frequency $\omega\ell/\bar{c}$ and its phase $\phi$, provided the structural properties $\lambda$ and $r$ are given. 

\begin{figure}[h!]
\includegraphics[width=1\linewidth]{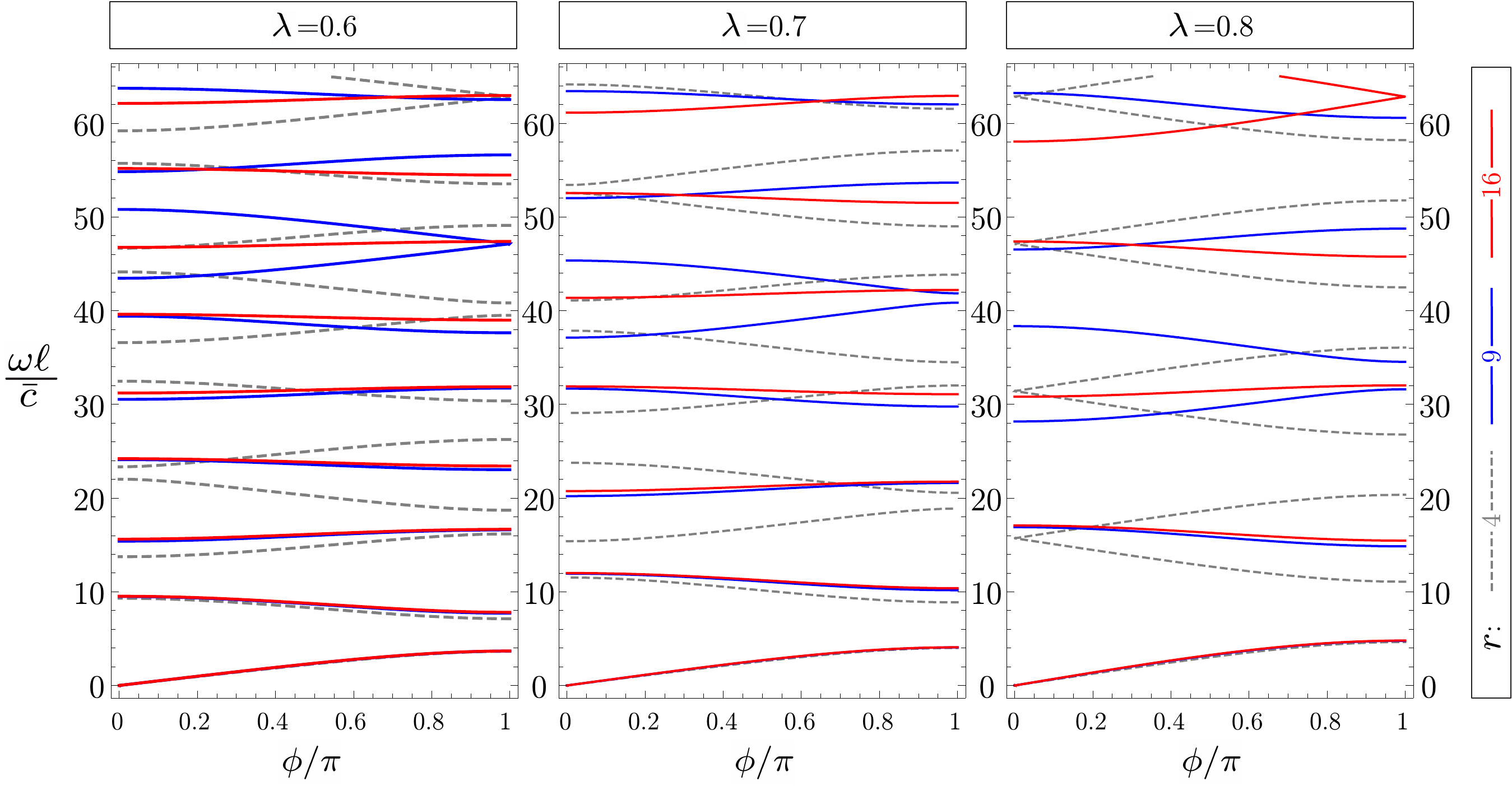}
	\caption{\label{fig:disp-diag-nbw} Dispersion diagrams $\omega\ell/\bar{c}$ - $\phi/\pi$  for periodic structures differing in the value of $r=\sqrt{E_B/E_A}=\{4,9,16\}$ (and respectively reported as dashed grey, continuous blue, and continuous red lines) and with $\lambda=\{0.6,0.7,0.8\}$, increasing from left to right.  The stop band widths increase with the increase of $r$ and of $\lambda$.}
\end{figure}

The dispersion diagrams  $\omega\ell/\bar{c}$--$\phi$ reported in Fig. \ref{fig:disp-diag-nbw} for System \textbf{(0)}  with $\lambda=\{0.6,$ $0.7,$ $0.8\}$ and $r=\{4,9,16\}$ are well-known \cite{movchan2002asymptotic} and show the presence of pass bands and stop bands (also called \lq band gaps'), namely, 
intervals of the frequency ratio $\omega\ell/\bar{c}$ inside which a Bloch-wave may and may not propagate, respectively. 

It is remarked that these pass and stop bands, only exist for System \textbf{(0)} (not involving transverse displacements). Flexural vibrations 
allow the possibility of axial vibrations for Systems \textbf{(I)} and \textbf{(II)} even for angular frequencies $\omega$ inside the stop bands defined for 
System \textbf{(0)}. However the pass and stop band structure of the underlying System \textbf{(0)} is 
related to interesting features for Systems \textbf{(I)} and \textbf{(II)}, as shown later in this Section.

\subsection{Flexural vibrations with configurational forces acting  on one substructure: System \textbf{(I)}}

In System \textbf{(I)} transverse oscillations may only exist within the substructure $B$, so that the parameter $\lambda$ is unconstrained (namely, eqn (\ref{length_constraint}) does not hold). 
In this case only configurational forces at the end of the sliding sleeve exerted on the substructure $B$ are generated. 
Considering that the substructure $B$   oscillates transversally under  mode $n_B$, the angular frequency $\omega$ can be obtained from eqn (\ref{betavalues}) as
\begin{equation}
\omega=\frac{2 c_B \mathcal{R}_B \left(\Xi^{(n_B)}\right)^2}{(\lambda-2\delta)^2\ell^2},
\end{equation} 
which, under the assumptions (\ref{eq:simple-par}), (\ref{eq:par1-I}), and (\ref{eq:par2-I}), reduces to the following dimensionless form
\begin{equation}\label{omegaI}
\frac{\omega(n_B)\ell}{\overline{c}}=\frac{2 r}{ \Lambda} \left(\frac{\Xi^{(n_B)}}{\lambda-2\delta}\right)^2,
\end{equation} 
where $\Lambda=\ell\sqrt{\overline{\mathcal{S}}/\overline{\mathcal{I}}}$ is the slenderness of the periodic structural cell.
The dimensionless frequency $\omega\ell/\overline{c}$, eqn (\ref{omegaI}), defines the oscillations of System \textbf{(I)} for given values of $r$, $\lambda$, $\delta$, $\Lambda$, and mode number $n_B$, 
whose longitudinal amplitudes  can be obtained from the solution of the linear system (\ref{eq:kernel00}), once a non-null transverse  amplitude $\textsf{V}_{BC1}^{[0]}$ is imposed (and by assuming $\textsf{V}_{AC1}^{[0]}=0$).
Because such an evaluation requires the inversion of the matrix $\mathbb{M}$, two cases may be distinguished:
\begin{itemize}
\item the dimensionless frequency $\omega\ell/\overline{c}$ belongs to one of the band gaps of System \textbf{(0)}. In this case, the  matrix $\mathbb{M}$ is always non-singular so that finite longitudinal amplitudes are always attained for a given finite transverse amplitude $\textsf{V}_{BC1}^{[0]}$;

\item the dimensionless frequency $\omega\ell/\overline{c}$ 
does not belong to a band gap. In this case, the  matrix $\mathbb{M}$ becomes  singular for values of $\omega\ell/\overline{c}$ and $\phi$ lying on the dispersion curve, for which the  \textit{resonance condition}, eqn (\ref{annihilation_determinante12x12}), is attained. It follows that oscillations defined by pairs of $\omega\ell/\overline{c}$ and $\phi$ close to  the dispersion curve are governed by 
large longitudinal amplitudes (approaching infinity)  at given finite transverse amplitude $\textsf{V}_{BC1}^{[0]}$.
\end{itemize} 

It follows from the above discussion that longitudinal vibrations are generated when transverse vibrations are present (in Systems \textbf{(I)} and \textbf{(II)}). For this reason, such longitudinal vibrations  are referred to as nested Bloch waves, as they are 'nested' to the transverse ones. Indeed, while the classical Bloch longitudinal waves  (in System \textbf{(0)}) have frequency defined by eqn (\ref{detridotto}) and undefined amplitude, the nested Bloch longitudinal waves have frequency defined by relation (\ref{eq:long-freq}) and amplitude defined in relation of the transverse oscillations amplitudes $\textsf{V}_{AC1}^{[0]}$ and $\textsf{V}_{BC1}^{[0]}$ via the linear system (\ref{eq:kernel00}).

It is also worth to remark that in the presence of transverse vibrations, longitudinal vibrations may become possible at frequencies that are forbidden in the absence of flexural motion.

\subsection{Flexural vibrations  with configurational forces acting  on two substructures: System \textbf{(II)}}

The presence of transverse vibrations within both substructures 
$A$ and $B$ subject to configurational forces 
is possible only when $\lambda$ satisfies eqn (\ref{length_constraint}). Considering oscillations under modes $n_A$ and $n_B$ within substructures $A$ and $B$, under the assumptions described by eqns (\ref{eq:simple-par}), (\ref{eq:par1-I}), and (\ref{eq:par2-I}), the parameter $\lambda$ is given by
\begin{equation}\label{eq:par2-I}
\lambda=2 \delta+\frac{1-4\delta}
{1+\ds\frac{\Xi^{(n_A)}}{\sqrt{r}\,\Xi^{(n_B)}}},
\end{equation}
so that 
using eqn (\ref{eq:par2-I}) the dimensionless angular frequency (\ref{omegaI}) reduces to
\begin{equation}\label{omegaII}
\frac{\omega(n_A,n_B)\ell}{\overline{c}}=\frac{2}{ \Lambda} \left(\frac
{\Xi^{(n_A)}+\sqrt{r} \,\Xi^{(n_B)}}{1-4\delta}
\right)^2.
\end{equation} 
Similarly to System \textbf{(I)} the dimensionless frequency $\omega\ell/\overline{c}$, eqn (\ref{omegaII}), is defined once the three parameters  
$r$, $\Lambda$, $\delta$, plus the modes numbers $n_A$ and $n_B$, are given. 
Only 
parameters $r$, $\delta$, $\Lambda$,  plus the mode number $n_B$, have to be prescribed for Systems \textbf{(I)} and \textbf{(II)}. 
The mode number $n_A$ is not present in System \textbf{(I)} and is replaced there by the parameter $\lambda$ (which is constrained in System \textbf{(II)} by eqn (\ref{eq:par2-I}), but remains free in System \textbf{(I)}).

In the particular case when the two substructures  oscillate transversally under the same mode $n_A=n_B=n$, the relations (\ref{eq:par2-I}) and (\ref{omegaII}) simplify as
\begin{equation}\label{lambda2fin}
\lambda=2 \delta+\frac{1-4\delta}
{1+\ds\frac{1}{\sqrt{r}}} ,\qquad
\frac{\omega(n)\ell}{\overline{c}}=\frac{2}{\Lambda} \left(\frac
{1+\sqrt{r}}{1-4\delta} \right)^2 {\Xi^{(n)}}^2.
\end{equation}

\begin{figure}[p]
\begin{center}
\includegraphics[width=0.9\linewidth]{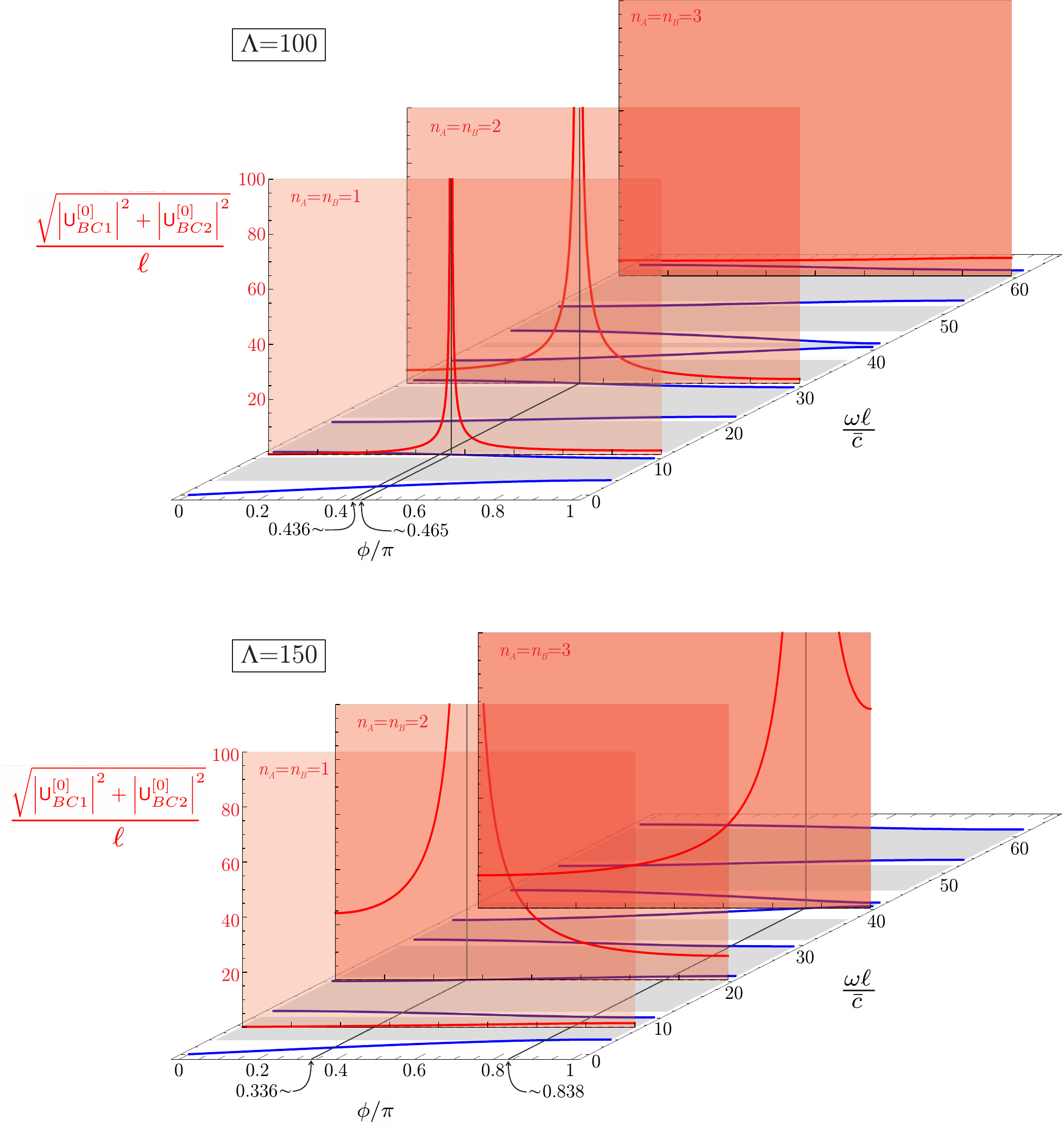}
\end{center}

	\caption{\label{fig:resonances} Dimensionless amplitude modulus ($\sqrt{\left|\textsf{U}_{BC1}^{[0]}\right|^2+\left|\textsf{U}_{BC2}^{[0]}\right|^2}/\ell$, red curves) of the longitudinal nested oscillation in the central part of the substructure $B$ realized by a transverse oscillation  under modes $n_A=n_B=\{1,2,3\}$  with amplitude defined by 
				$\textsf{V}_{AC1}^{[0]}/\ell=\textsf{V}_{BC1}^{[0]}/\ell=1$ on System \textbf{(II)} with $r=9$, $\delta =0.05$ and $\lambda=0.7$.
	The horizontal axis reports the phase angle $\phi$, while 
	the out of plane axis the dimensionless frequency 
				$\omega\ell/\overline{c}$, so that the blue curves are the dispersion diagram for purely longitudinal oscillations (shaded light blue regions represent stop bands for classical Bloch waves). 
			Resonance occurs for $\Lambda = 100$ ($\Lambda = 150$) at $\phi\approx \{0.436, 0.465\}\pi$ ($\phi\approx \{0.336, 0.838\}\pi$) under first and second (second and third) modes.}
\end{figure}

The dynamic response of System \textbf{(II)} is depicted in Fig. \ref{fig:resonances} through a three-dimensional representation in which the longitudinal amplitude modulus of substructure $B$, given by $\sqrt{\left|\textsf{U}_{BC1}^{[0]}\right|^2+\left|\textsf{U}_{BC2}^{[0]}\right|^2}/\ell$ (vertical axis), is reported with  
varying the phase angle $\phi$ (horizontal axis) and the dispersion curve at different values of dimensionless frequency $\omega\ell/\overline{c}$ (out of plane axis). 
The system is considered to have both substructures  oscillating transversally under the same mode, $n_A=n_B=n$ with $n=\{1,2,3\}$, with transverse amplitude defined by $\textsf{V}_{AC1}^{[0]}/\ell=\textsf{V}_{BC1}^{[0]}/\ell=1$. System parameters are selected  as $r = 9$ and $\delta= 0.05$, to which $\lambda = 0.7$ corresponds from eqn. (\ref{lambda2fin}). The behaviour is shown for two values of structural cell slenderness, $\Lambda=100$ (upper part) and $\Lambda=150$ (lower part), for which the first three modes correspond, respectively, to the dimensionless frequencies $\omega\ell/\overline{c}\approx\{11.186, 30.843, 60.451\}$ and $\omega\ell/\overline{c}\approx\{7.458, 20.562, 40.300\}$.

It can be noted from the figure that for the structure with $\Lambda=100$ ($\Lambda=150$) resonance occurs under modes $n=\{1,2\}$ ($n=\{2,3\}$) at respectively phase angles 
$\phi\approx \{0.465, 0.436\}\pi$ ($\phi\approx \{0.336, 0.838\}\pi$), where the dispersion curve is intersected. Resonance does not occur at any phase angle $\phi$ under third (first) mode for $\Lambda=100$ ($\Lambda=150$), for which there is no intersection with the dispersion curve.

\section{Conclusions}
The  analysis of a periodic system composed of elastic beams and sliding sleeves shows the important role played by configurational forces in dynamics.
Configurational forces introduce a non-linearity, which couples transverse motion to longitudinal one, and provide nesting  of 
axial vibrations to the flexural ones. The 
 band gap structure defined by purely axial vibration is broken and the propagation of nested longitudinal waves may become possible at frequencies forbidden in the absence of transverse vibrations.
 The flexural vibration is also shown
to generate  resonance for the nested longitudinal waves at frequencies within the pass band of classical Bloch waves. These results suggest applications of configurational forces in the design of dynamic sensors.

\vspace*{5mm} \noindent {\sl Authors' Contributions.} FDC, DT, DB performed the calculations and computations.  FDC, DB, NVM, ABM planned the research. FDC, DT, DB wrote the paper. All authors read and approved the manuscript.

\vspace*{5mm} \noindent {\sl Acknowledgments.} FDC gratefully acknowledges financial support from the European Union's
Horizon 2020 research and innovation programme under the
Marie Sklodowska-Curie grant agreement \lq
INSPIRE - Innovative ground interface concepts for structure protection'  PITN-GA-2019-813424-INSPIRE.
DT, NVM, and ABM gratefully
acknowledge financial support from the grant ERC Advanced Grant \lq
Instabilities and nonlocal multiscale modelling of materials' ERC-2013-ADG-340561-INSTABILITIES. DB gratefully
acknowledges financial support from PRIN 2015 \lq Multi-scale mechanical models for the design and optimization of micro-structured smart materials and metamaterials' 2015LYYXA8-006. The authors also acknowledge funding from the Italian Ministry of Education, University and Research (MIUR) in the frame of the "Departments of Excellence" grant L. 232/2016.


\vspace*{10mm}



\begin{thebibliography}{99}



\bibitem{armanini2019}
Armanini A, Dal Corso F, Misseroni D, Bigoni D 2019.
Configurational forces and nonlinear structural dynamics.
\emph{J. Mech. Phys. Sol.}  130, 82--100

\bibitem{baci2017}
Bacigalupo A, Gambarotta L 2017. Wave propagation in non-centrosymmetric beam-lattices with lumped masses: Discrete and micropolar modeling. \emph{Int. J. Sol. Struct. } 118--119, 128--145

\bibitem{baci2018}
Bacigalupo A, Lepidi M 2018.
Acoustic wave polarization and energy flow in periodic beam lattice materials. \emph{Int. J. Sol. Struct. } 147, 183--203

\bibitem{balabukh1970work}
Balabukh LI, Vulfson MN, Mukoseev BV Panovko YaG 1970.
On work done by reaction forces of moving supports.
\emph{Research on Theory of Constructions}, 18, 190--200

\bibitem{ballarini} 
Ballarini R, Royer-Carfagni G 2016. A Newtonian interpretation of configurational forces on dislocations and cracks. \emph{J. Mech. Phys. Sol.} 95, 602-- 620 

\bibitem{Bigoni2014PRSA}
Bigoni D, Dal Corso F, Misseroni D, Bosi F 2014.
Torsional locomotion
\emph{Proc. R. Soc. A}, 470.2171, 20140599

\bibitem{Bigoni2014JMPS}
Bigoni D, Bosi F, Dal Corso F, Misseroni D 2014.
Instability of a penetrating blade
\emph{J. Mech. Phys. Sol.} 64, 411--425

\bibitem{Bigoni2015MM}
Bigoni D, Dal Corso F, Bosi F, Misseroni D 2015.
Eshelby-like forces acting on elastic structures: theoretical and experimental proof.
\emph{Mech. Mater.}, 80, 368--374

\bibitem{bigonigei}
Bigoni D, Gei M, Movchan AB 2008. 	
Dynamics of a prestressed stiff layer on an elastic half space: filtering and band gap characteristics of periodic structural models derived from long-wave asymptotics 	\emph{J. Mech. Phys. Sol.} 56(7), 2494--2520 

\bibitem{bosiarmscale}
Bosi F, Misseroni D, Dal Corso F, Bigoni D. 2014.
An elastica arm scale.
\emph{Proc. R. Soc. A} 470:20160870

\bibitem{bosidripping}
Bosi F, Misseroni D, Dal Corso F, Bigoni D 2015.
Self-encapsulation, or the \lq dripping' of an elastic rod.
\emph{Proc. R. Soc. A}, 471: 20150195

\bibitem{bosiinjection} 
Bosi F, Misseroni D, Dal Corso F, Bigoni D 2015.
Development of configurational forces during the injection of an elastic rod. \emph{Ext. Mech. Lett.}, 471 83--88

\bibitem{bosi2016asymptotic}
Bosi F, Misseroni D, Dal Corso F, Neukirch S, Bigoni D 2016.
Asymptotic self-restabilization of a continuous elastic structure.
\emph{Phys. Rev. E}, 94 (6): 063005

\bibitem{brun}
Brun M, Movchan AB, Slepyan LI 2013. Transition wave in a supported heavy beam. \emph{J. Mech. Phys. Sol.} 61(10), 2067--2085


\bibitem{carta0}
Carta G,  Brun M 2015. Bloch-Floquet waves in flexural systems with continuous and discrete element. \emph{Mech. Mat.} 87, 11--26

\bibitem{carta}
Carta G,  Jones IS, Movchan NV, Movchan AB,  Nieves MJ 2017. Gyro-elastic beams for the vibration reduction of long flexural systems. \emph{Proc. R. Soc. A} 473, 20170136

\bibitem{carta2} 
Carta G, Nieves MJ, Jones IS, Movchan NV, Movchan AB 2018. Elastic chiral waveguides with gyro-hinges, \emph{Quart. J. Mech. Appl. Math.} 71, 157--185

\bibitem{casadei} 
Casadei F, Bertoldi K 2014.
Wave propagation in beams with periodic arrays of airfoil-shapedresonating units, \emph{J. Sound  Vibr.}  333 (24), 6532--6547


\bibitem{dal2017serpentine}
Dal Corso F, Misseroni D, Pugno NM, Movchan AB, Movchan NV, Bigoni D 2017.
Serpentine locomotion through elastic energy release.
\emph{J. R. Soc. Interface}, 14: 20170055

\bibitem{deng}
Deng B, Wang P, He Q, Tournat V, Bertoldi K 2018. Metamaterials with amplitude gaps for elastic solitons
\emph{Nature Comm.}, 9, 3410


\bibitem{eshelby1}
Eshelby JD 1951.
The force on an elastic singularity.
\emph{Phil. Trans. R. Soc. A}, 244--877, 87--112

\bibitem{eshelby2}
Eshelby JD 1956.
The continuum theory of lattice defects.
\emph{Solid State Phys.}, 3.C, 79--144

\bibitem{eshelby3}
Eshelby JD 1970.
Energy relations and the energy-momentum tensor in continuum mechanics.
\emph{Inelastic Behaviour of Solids}, 17--115

\bibitem{fraternali}
Fraternali F, Amendola A 2017.
Mechanical modeling of innovative metamaterials alternating pentamode lattices and confinement plates.
\emph{J. Mech. Phys. Sol.} 99, 259--271



\bibitem{garau}
Garau M, Carta G, Nieves MJ, Jones IS, Movchan NV, Movchan AB 2018. Interfacial waveforms in chiral lattices with gyroscopic spinners. \emph{Proc. R. Soc. A} 474: 20180132

\bibitem{hanna} 
Hanna JA, Singh H, Virga EG 2018. Partial Constraint Singularities in Elastic Rods. \emph{J. Elas.} 133(1), 105--118 

\bibitem{liakou1} 
Liakou A 2018. Constrained buckling of spatial elastica: Application of optimal control method. \emph{J. App. Mech. ASME} 85(8),081005

\bibitem{liakou2} 
Liakou A  2018. Application of optimal control method in buckling analysis of constrained elastica problems. \emph{Int. J. Sol. Struct.} 141-142, 158--172

\bibitem{liakou3} 
Liakou A, Detournay E 2018. Constrained buckling of variable length elastica: Solution by geometrical segmentation. \emph{Int. J. Non-Linear Mech.} 99,  204--217.


\bibitem{Maurin2014_WM}
Maurin FPR, Spadoni A 2014.
Low-frequency wave propagation in post-buckled structures. 
\emph{Wave Motion} 51, 323--334

\bibitem{Maurin2014_JSV}
Maurin FPR, Spadoni A 2014.
Wave dispersion in post-buckled structures. 
\emph{J. Sound Vibr.}, 333,
4562--4578

\bibitem{Maurin2016_WM1}
Maurin FPR, Spadoni A 2016.
Wave propagation in periodic buckled beams. Part I: Analytical models and numerical simulations,
\emph{Wave Motion} 66, 190--209

\bibitem{Maurin2016_WM2}
Maurin FPR, Spadoni A 2016.  Wave propagation in periodic buckled beams. Part II: Experiments.
\emph{Wave Motion}  66, 210--219


\bibitem{morinigei}
Morini L, Gei M 	2018. 	Waves in one-dimensional quasicrystalline structures: dynamical trace mapping, scaling and self-similarity of the spectrum. \emph{J. Mech. Phys. Sol.}
119, 83--103 


\bibitem{movchan2002asymptotic}
Movchan AB, Movchan NV, Poulton CG 2002. Asymptotic models of fields in dilute and densely packed composites. World Scientific Publ.

\bibitem{Nadkarni2014}
Nadkarni N, Daraio C,  Kochmann DM 2014.
Dynamics of periodic mechanical structures containing bistable elastic elements: From elastic to solitary wave propagation
\emph{Phys. Rev. E} 90, 023204

\bibitem{Nadkarni2016}
Nadkarni N, Arrieta AF, Chong C, Kochmann DM, Daraio C 2016. Unidirectional
transition waves in bistable lattices. \emph{Phys. Rev. Lett.}, 116, 244501


\bibitem{nievesIJSS} 
Nieves MJ, Mishuris GS, Slepyan LI 2017. Transient wave in a transformable periodic flexural structure. \emph{Int. J. Sol. Struct.} 112, 185--208

\bibitem{nievesJMPS} 
Nieves MJ, Carta G, Jones IS, Movchan AB,  Movchan NV 2018. Vibrations and elastic waves in chiral multi-structures. \emph{J. Mech. Phys. Solids} 121, 387--408

\bibitem{oreilly} 
O'Reilly OM 2015. 	Some perspectives on Eshelby-like forces in the elastica arm scale. \emph{Proc. R. Soc. A} 471(2174),20140785

\bibitem{oreilly2} 
O'Reilly, OM 2017. \emph{Modeling Nonlinear Problems in the Mechanics of Strings and Rods:
The Role of the Balance Laws}. Springer

\bibitem{singh} 
Singh H, Hanna JA 2019. On the Planar Elastica, Stress, and Material Stress. \emph{J. Elas.} https://doi.org/10.1007/s10659-018-9690-5

\bibitem{tallarico2017tilted}
Tallarico D, Movchan NV, Movchan AB, Colquitt DJ 2017. Tilted resonators in a triangular elastic lattice: chirality, Bloch waves and negative refraction. \emph{J. Mech. Phys. Sol.}, 103, 236--256


\bibitem{tallarico2017edge}
Tallarico D, Trevisan A, Movchan NV, Movchan AB 2017.
Edge waves and localization in lattices containing tilted resonators. \emph{Front. Mat.}, 4, 16

\bibitem{tallarico2017propagation}
Tallarico D, Movchan NV, Movchan AB,  Camposaragna M 2017. Propagation and filtering of elastic and electromagnetic waves in piezoelectric composite structures,
\emph{Math. Meth. Appl. Sci.},
 40 (9), 3202--3220

\end{thebibliography}
\end{document}